%% file: main.tex
\documentclass[12pt]{article}

\usepackage[letterpaper, margin=1in]{geometry}

\usepackage{graphicx}
\usepackage[affil-it]{authblk}
\usepackage{caption}
\usepackage{subcaption}
\usepackage{siunitx}
\usepackage{tikz}
\usepackage{pgfplots}
\usepackage{verbatim}
\usepackage{amsmath}
\usepackage{pgfplotstable}

\usetikzlibrary{spy,backgrounds}

\usepackage{hyperref}
\hypersetup{
    colorlinks=true,
    linkcolor=blue,
    filecolor=magenta,      
    urlcolor=blue,
}

\begin{document}

\title{New smoothed particle hydrodynamics (SPH) formulation for modeling heat conduction with solidification and melting\\}
\author[1]{Amirsaman Farrokhpanah\thanks{farrokh@mie.utoronto.ca}}
\author[1]{Markus Bussmann}
\author[1]{Javad Mostaghimi}
\affil[1]{\normalsize Department of Mechanical \& Industrial Engineering, University of Toronto}
\date{April 2017}

Published in: \href{http://www.tandfonline.com/doi/full/10.1080/10407790.2017.1293972}{Numerical Heat Transfer, Part B: Fundamentals}, 71(4), pp. 299-312

DOI: \href{https://doi.org/10.1080/10407790.2017.1293972}{10.1080/10407790.2017.1293972}

Direct Access: \href{http://www.tandfonline.com/eprint/eYVQNFNMFYdekIfWPakm/full}{Full Text}

\let\newpage\relax
\maketitle

\begin{abstract}
\noindent When modelling phase change, the latent heat released (absorbed) during solidification (melting) must be included in the heat transfer equation. In this paper, different SPH methods for the implementation of latent heat, in the context of transient heat conduction, are derived and tested. First, SPH discretizations of two finite element methods are presented, but these prove to be computationally expensive. Then, by starting from a simple approximation and enhancing accuracy using different numerical treatments, a new SPH method is introduced, that is fast and easy to implement. An evaluation of this new method on various analytical and numerical results confirms its accuracy and robustness.
\end{abstract}


\section{Introduction}
\label{Introduction}

Heat transfer and phase change are of importance in many engineering applications, e.g. coating surfaces with paint or metal, freezing and thawing of food products, and the casting of plastics and metals. There are a wide range of methods available for numerical prediction of transient phase change problems. Knowledge of the solid-liquid front position is of key importance. In many numerical methods for phase change, the effect of latent heat is added as a source term to the heat transfer equation, like the methods of Passandideh Fard \cite{Pasandideh_96_2} and Voller \cite{Voller_87}, or by modifying the heat capacity coefficient, as in Thomas et al. \cite{Thomas_84}, Hsiao \cite{Hsiao_86}, and Dalhuijsen et al. \cite{Dalhuijsen_86}.

The finite volume method was first used for the study of phase change heat transfer, followed by finite element methods. This transition was due to the flexibility of the finite element method for complex boundary conditions and implementations \cite{Dalhuijsen_86}. In this paper, a Smoothed Particle Hydrodynamics method (SPH), or in a more general sense, integral interpolations are investigated for phase change processes. SPH is a Lagrangian mesh-free CFD method introduced in 1977 by Lucy \cite{Lucy_97} and Gingold and Monaghan \cite{Gingold_77}, and was initially applied to astrophysics. Since 1977, many studies have been conducted on the accuracy and applicability of SPH to various fluid problems \cite{Morris_97}. Tsunami simulations \cite{Liu_08}, simulations of floating bodies like ships \cite{Cartwright_04}, and multiphase flow studies \cite{Hu_06, Grenier_08, Tartakovsky_09} are all examples of SPH applications. The main advantage of SPH is its ability to handle complex geometries. The implementation of different boundary conditions in SPH is also straightforward. Moreover, SPH codes are easily parallelized.

Several studies have attempted to solve phase change problems using SPH. Cleary et al. \cite{Cleary_98, Cleary_06} focused on the explicit inclusion of latent heat after solving the enthalpy form of the heat conduction equation. This method has been used to model casting \cite{Cleary_10} and solidification of molten metal drops impacting a surface \cite{Zhang_07,Zhang_08,Zhang_09}. Monaghan et al. \cite{Monaghan_05} modelled the solidification of pure and binary alloys, where phase change was simulated by removing liquid particles and transferring their mass to a stationary grid of solid particles. The drawback of this method is the need for virtual particles, which results in a doubling of the number of particles inside the domain, and makes the method computationally expensive and difficult to implement into available solvers.

In this study, a new SPH formulation, using integral interpolations is introduced for modelling transient heat conduction with phase change. Application of the method to various available analytical and numerical results demonstrates the robustness of the method. Compared to other effective heat capacity implementations, this method is fast and yet more accurate.

\section{Incorporation of Latent Heat into an Effective Heat Capacity}

\subsection{Governing Equations}

\input{ch_1}

\subsection{Numerical Implementations of Latent Heat Release}

\input{ch_2}

\section{Results and Discussion}

We now present the results of a series of tests of the five methods, referred to as equations \ref{eq:Cp_20}, \ref{eq:Cp_21}, \ref{eq:Cp_12}, \ref{eq:Cp_15}, and \ref{eq:Cp_18}.

\input{ch_3}

\clearpage

\section{Conclusions}

Modelling conduction heat transfer with phase change in SPH has been investigated. The release/absorption of latent heat during phase change is accounted for by modifying the heat capacity in the energy equation. A new approach is introduced, which uses smoothing and superposition of two kernels to gradually release/absorb the latent heat near the phase change temperature. Compared to a number of alternatives, this new approach yields accurate solutions while limiting the computational cost. The approximations proposed here, although based on a SPH formulation, are also applicable to grid-based methods.

\bibliographystyle{unsrt}
\bibliography{main}

\end{document}

%% file: ch_1.tex
Transient heat conduction is mathematically represented as
\begin{equation}\label{eq:Cp_7}
\frac{\partial H}{\partial t}=\nabla \cdot (k\nabla T)
\end{equation}
By expressing $H$ as a function of $T$, equation \ref{eq:Cp_7} can be written as

\begin{equation}\label{eq:Cp_8}
\frac{dH}{dT}\frac{\partial T}{\partial t}=\nabla \cdot (k  \nabla T)
\end{equation}
In this form, equation \ref{eq:Cp_8} does not account for the release (absorption) of latent heat during solidification (melting). To do that, the parameter

\begin{equation}\label{eq:Cp_9}
C=\frac{dH}{dT}
\end{equation}
known as the effective heat capacity, can be modified to include the effect of latent heat, as follows \cite{Bonacina_73, Hsiao_83}

\begin{equation}\label{eq:Cp_10}
C = \left\{ \begin{array}{lll}
        C_s & \mbox{$T<T_m$}\\
        C_m+L\delta(T-T_m) & \mbox{$T=T_m$}\\
        C_l & \mbox{$T>T_m$}\\
\end{array} \right. 
\end{equation} 
where $\delta(T-T_m)$ the Dirac delta function. It can be easily confirmed that using this form of $C$ yields a total heat release of $\int_{-\infty}^{\infty} L\delta(T-T_m)dT=L$.

%% file: ch_2.tex
First, SPH discretizations of two finite element methods for solving equation \ref{eq:Cp_7} are presented. Then, by starting with a simple approximation, new SPH interpolations for the inclusion of latent heat via equation \ref{eq:Cp_10} are derived. This leads to five different methods, that are presented in this paper as equations \ref{eq:Cp_20}, \ref{eq:Cp_21}, \ref{eq:Cp_12}, \ref{eq:Cp_15}, and \ref{eq:Cp_18}.

\subsubsection{Variational Methods}
A first approach is to use the variation of enthalpy as a function of temperature to calculate the value of C using equation \ref{eq:Cp_9}. The first equation presented here was developed by Del Diudice et al. \cite{Giudice_78} by transforming equation \ref{eq:Cp_9} using the direction of the temperature gradient, $s$, as \cite{Dalhuijsen_86}

\begin{equation}\label{eq:Cp_19}
 C = \frac{\partial H / \partial s}{\partial T / \partial s}=\frac{(\partial H / \partial x)l_{sx}+(\partial H/\partial y)l_{sy}+(\partial H/\partial z)l_{sz}}{\partial T / \partial s}
\end{equation} 
with direction cosines defined as $l_{s \alpha}=\frac{\partial T}{\partial \alpha}/\frac{\partial T}{\partial s}$ and $\frac{\partial T}{\partial s}=\sqrt{(\frac{\partial T}{\partial x})^2+(\frac{\partial T}{\partial y})^2+(\frac{\partial T}{\partial z})^2}$. This gives an effective heat capacity

\begin{equation}\label{eq:Cp_20}
 C =\frac{(\partial H / \partial x)(\partial T/\partial x)+(\partial H/\partial y)(\partial T/\partial y)+(\partial H/\partial z)(\partial T/\partial z)}{(\partial T/\partial x)^2+(\partial T/\partial y)^2+(\partial T/\partial z)^2}
\end{equation} 

Another approach is similar to that of Lemmon \cite{Lemmon_79}, where $s$ is taken as the normal direction of the interface, and yields

\begin{equation}\label{eq:Cp_21}
 C =\left[ \frac{(\partial H / \partial x)^2+(\partial H/\partial y)^2+(\partial H/\partial z)^2}{(\partial T/\partial x)^2+(\partial T/\partial y)^2+(\partial T/\partial z)^2}\right]^{1/2}
\end{equation} 

These two equations were implemented in a finite element context by Thomas et al. \cite{Thomas_84} and Dalhuijsen et al. \cite{Dalhuijsen_86}. To discretize equations \ref{eq:Cp_20} and \ref{eq:Cp_21} for SPH, an approach inspired by Del Giudice et al. \cite{Giudice_78} is used here. In their finite element method, the enthalpy is smoothed by a shape function before using the above equations. Here, the value of enthalpy for each SPH particle is calculated from its temperature using \cite{Dalhuijsen_86}

\begin{equation}\label{eq:Cp_22}
 C = \left\{ \begin{array}{lll}
        \int_{T_{ref}}^{T}\rho C_s(T) dT      & \mbox{$T < T_1$}\\
        \int_{T_{ref}}^{T_1}\rho C_s(T) dT    + \int_{T_{1}}^{T}\rho \left( dL/dT+C_m(T)\right) dT          & \mbox{$T_1 \leq T \leq T_2$}\\
        \int_{T_{ref}}^{T_1}\rho C_s(T) dT    +  \rho L + \int_{T_1}^{T_2}\rho C_m(T) dT + \int_{T_2}^{T}\rho C_l(T) dT & \mbox{$T > T_2$}\\
\end{array} \right. 
\end{equation}

To increase accuracy and stability, we propose that the value of enthalpy at each particle is smoothed using 

\begin{equation}\label{eq:Cp_23}
H_i^s=\sum_j \frac{m_j}{{\rho}_j} H_j W^P (\boldsymbol{x_j} - \boldsymbol{x_i},h)
\end{equation} 
With the value of enthalpy known, the gradients of enthalpy that appear in equations \ref{eq:Cp_20} and \ref{eq:Cp_21} can be calculated using 

\begin{equation}\label{eq:Cp_24}
\left( \nabla H \right)_i =\sum_j \frac{m_j}{{\rho}_j} \left[ H_j^s - H_i^s \right] \boldsymbol{\nabla W^P} (\boldsymbol{x_j} - \boldsymbol{x_i},h)
\end{equation}

A similar formulation is used to calculate the first gradient of temperature. The second gradient of temperature that appears in these equations is calculated by taking the gradient of the first gradient:

\begin{equation}\label{eq:Cp_25}
\left( \nabla ^2 T \right)_i = \rho_i\sum_j m_j \left[ \frac{\boldsymbol{\nabla T}_j}{\rho^2_j}+\frac{\boldsymbol{\nabla T}_i}{\rho^2_i} \right] \cdot \boldsymbol{\nabla W^P} (\boldsymbol{x_j} - \boldsymbol{x_i},h)
\end{equation}

\subsubsection{Step Release}
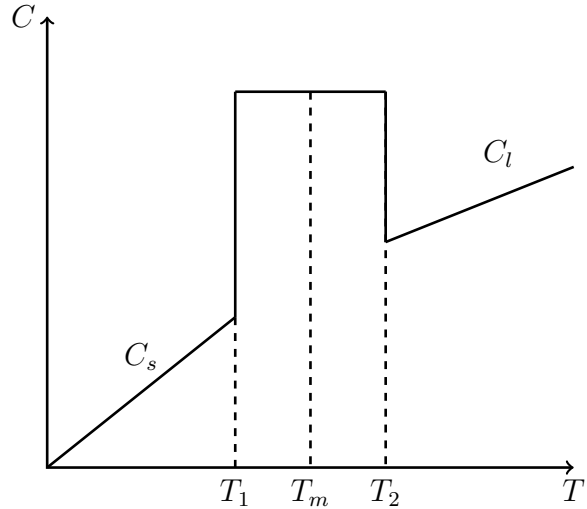
\begin{figure}[h]
        \centering
\begin{tikzpicture}
    \draw [<->,line width=1pt] (0,6) node (yaxis) [left] {$C$}  |- (7,0) node (xaxis) [below] {$T$};

    \draw[line width=1pt] (0,0)  -- (2.5,2) coordinate (a_2);
    \draw[line width=1pt] (a_2) -- (2.5,5) coordinate (a_3);
    \draw[line width=1pt] (a_3) -- (4.5,5) coordinate (a_4);
    \draw[line width=1pt] (a_4) -- (4.5,3) coordinate (a_5);
    \draw[line width=1pt] (a_5) -- (7,4) coordinate (a_6);

    \draw[dashed,line width=1pt]   (a_2) -- (2.5,0) node[below] {$T_1$};
    \draw[dashed,line width=1pt]   (a_4) -- (4.5,0) node[below] {$T_2$};
    \draw[dashed,line width=1pt]   (3.5,5) -- (3.5,0) node[below] {$T_m$};

    \draw (1.25,1.8) node[below] {$C_s$};
    \draw (6,4.5) node[below] {$C_l$};

\end{tikzpicture}
\caption{Apparent heat capacity with a step jump for the latent heat.}
\label{fig:Cp_1}
\end{figure}

Calculation of the partial derivatives appearing in equations \ref{eq:Cp_20} and \ref{eq:Cp_21} demands many computational steps, and so motivates the search for alternative methods. Returning to equation \ref{eq:Cp_10}, it is convenient to replace $\delta$, that releases latent heat at a single temperature $T_m$, with a function that releases the latent heat over a finite temperature interval around $T_m$. Hsiao \cite{Hsiao_86} shows that a very simple substitute for $\delta$ is of the form

\begin{equation}\label{eq:Cp_11}
 C = \left\{ \begin{array}{lll}
        C_s & \mbox{$T<T_m-\Delta T$}\\
        C_m+\frac{L}{2\Delta T} & \mbox{$T_m - \Delta T \leq T \leq T_m + \Delta T$}\\
        C_l & \mbox{$T>T_m+\Delta T$}\\
\end{array} \right. 
\end{equation} 
Here the total latent heat released over the interval of $2\Delta T$ is still $L$. Figure \ref{fig:Cp_1} shows this step jump in the apparent heat capacity. 

The main drawback of this approach is the need for a phase change temperature interval. For some particles during a simulation, the value of temperature may jump from a value above this interval to a value below it in one time step \cite{Hsiao_86}. This means that the effect of latent heat might be bypassed if the temperature variation in a time step is large compared to $\Delta T$, and imposes a limitation on the solution time step \cite{Dalhuijsen_86}. Additionally, for the case of pure materials where solidification/melting occurs at a single temperature, the artificial phase change interval \cite{Thomas_84} has no physical meaning. A poor choice of this interval may cause solution inaccuracy.

For the phase change of alloys with a mushy zone, where solidification/melting occurs over a temperature range, equation \ref{eq:Cp_11} becomes

\begin{equation}\label{eq:Cp_12}
 C = \left\{ \begin{array}{lll}
        C_s & \mbox{$T \leq T_1$}\\
        C_m+\frac{L}{T_2-T_1} & \mbox{$T_1<T<T_2$}\\
        C_l & \mbox{$T \geq T_2$}\\
\end{array} \right. 
\end{equation}

\subsubsection{Gradual Release}

A second approach is to replace the Dirac delta function $\delta$ by an even smoothing function $W$ that has the identity property of $\delta$, $\int_{-\infty}^{\infty} W(T-T_m,h^*)dT=1$, and is defined over a temperature range, shown here as $\Delta T=k^* \times h^*$. The integer $k^*$ and the float $h^*$ are similar in definition to the usual SPH notations of $k$ and $h$, the only difference being that $k$ and $h$ are defined in the space domain and are based on particle positioning, while $k^*$ and $h^*$ are defined over the temperature domain of particles. In other words, where the value of $k \times h$ determines a radius around each particle within which its neighbours contribute to a heat transfer calculation, similarly, $k^* \times h^*$ defines a radius in the one dimensional domain of temperature around the melting temperature. This new radius acts as a second filter in addition to $k \times h$. In formulations where both filters are present, the only particles that contribute to heat transfer are those close to a certain particle in both the space and temperature domains.

\begin{figure}[h]
        \centering
\begin{tikzpicture}
    \draw [<->,line width=1pt] (0,5) node (yaxis) [left]{$W$}  |- (7,0) node (xaxis) [below] {$T$};
    \draw[line width=1pt,domain=1:6,smooth,variable=\x] plot ({\x},{ 3*exp(-(\x-3.5)^2) });
    \draw (5.5,3) node[below] {$W(T-T_m, h^*)$};
    \draw[<->,line width=1pt] (3.5, -0.7) -- (6, -0.7) node[pos=0.5,below]{$k^*h^*$};
    \draw[<-,line width=1.5pt]   (3.5,4) node[above] {$\delta (T-T_m)$} -- (3.5,0) node[below] {$T_m$};
\end{tikzpicture}
\caption{The smoothing function W.}
\label{fig:Cp_2}
\end{figure}
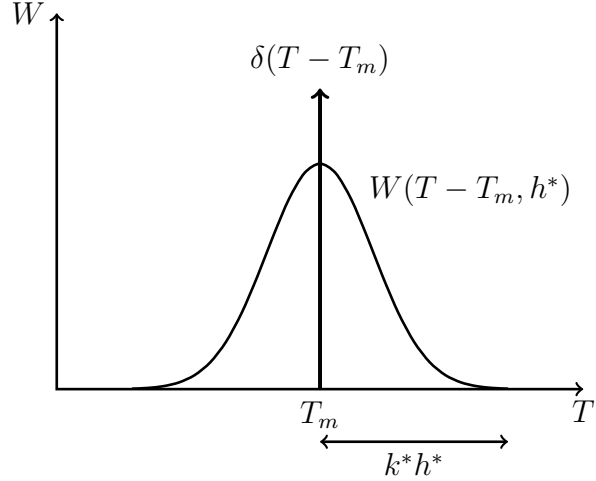

Replacing $\delta$ with the smoothing function $W$, depending on the choice of $W$, has been shown to have a second or higher order of accuracy \cite{Liu_03}. Figure \ref{fig:Cp_2} demonstrates the smoothing function. More detailed discussions on the derivation of a SPH formulation, and more general integral interpolations, can be found in \cite{Farrokhpanah_15, Farrokhpanah_12}. Using this definition, equation \ref{eq:Cp_10} can be rewritten as

\begin{equation}\label{eq:Cp_15}
 C = \left\{ \begin{array}{lll}
        C_s & \mbox{$T<T_m-\Delta T$}\\
        C_m+ LW^T(T-T_m,h^*) & \mbox{$T_m - \Delta T \leq T \leq T_m + \Delta T$}\\ 
        C_l & \mbox{$T>T_m+\Delta T$}\\
\end{array} \right. 
\end{equation}
where $W^T$ is a 1D kernel defined in the temperature domain.

\subsubsection{Smoothed Gradual Release}

The accuracy and efficiency of equation \ref{eq:Cp_15} can be further improved by smoothing the term $L W^T (T-T_m,h^* )$ with a higher order kernel \cite{Aleinov_95,Rudman_98}. This is achieved using the standard SPH interpolation \cite{Morris_00}

\begin{equation}\label{eq:Cp_6}
f(x_0)\cong -\sum_{j=1}^{N}\frac{m_j}{\rho_j}f(x_j)W(x_i-x_0,h)
\end{equation}

Summation here is performed over all $j$ particles located in the neighbourhood radius of particle $i$. With this definition, the latent heat in equation \ref{eq:Cp_15} at the position of an arbitrary particle $i$ is smoothed using

\begin{equation}\label{eq:Cp_17}
L_i^s =\sum_{j=1}^{N}\frac{m_j}{\rho_j}[LW^T(T_j-T_m,h^*)]W^P(\boldsymbol{x_j}-\boldsymbol{x_i},h)
\end{equation}
and hence equation \ref{eq:Cp_10} can be rewritten as

\begin{equation}\label{eq:Cp_18}
 C = \left\{ \begin{array}{lll}
        C_s & \mbox{$T<T_m-\Delta T$}\\
        C_m+ L^s & \mbox{$T_m - \Delta T \leq T \leq T_m + \Delta T$}\\
        C_l & \mbox{$T>T_m+\Delta T$}\\
\end{array} \right. 
\end{equation}
$\Delta T$ is identical to the radius of influence of the smoothing function, $W^T(T-T_m,h^*)$. In this formulation, the latent heat is distributed to particles based on the difference between the particle temperature and the melting temperature, and also based on how far other particles in its neighbourhood are from the reference melting temperature. In other words, this equation accounts for two variations: $1)$ changes in temperature of each particle with respect to the melting point, and $2)$ the variation of temperature in the domain with respect to the position vector $\boldsymbol{x}$. The term $LW^T(T_j-T_m,h^*)$ is a discretization of $L \delta$ and a function of temperature only. Furthermore, temperature itself is a function of particle position in the domain $T=T(x)$. The second term, $W^P(\boldsymbol{x_j}-\boldsymbol{x_i},h)$, accounts for the change of temperature with respect to position. The smoothing behaviour of equation \ref{eq:Cp_17} assures there is no sudden jump in enthalpy within the domain with respect to changes in the position vector $\boldsymbol{x}$.

%% file: ch_3.tex
\subsection{Two-Dimensional Problem: Phase Change in the Corner of a Square}
\label{sec:analyticalcomparison}

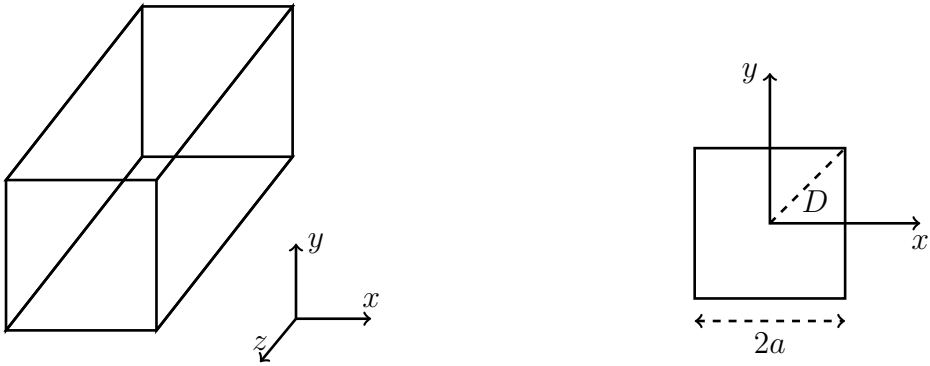
\begin{figure}[h]
	\centering
	\begin{subfigure}[b]{0.49\textwidth}
		\centering
		\begin{tikzpicture}
			\draw[->,line width=1pt] (xyz cs:x=2.2,y=-2,z=0) -- (xyz cs:x=3.2,y=-2,z=0) node[above] {$x$};
			\draw[->,line width=1pt] (xyz cs:x=2.2,y=-2,z=0)  -- (xyz cs:x=2.2,y=-1,z=0)  node[right]   {$y$};
			\draw[->,line width=1pt] (xyz cs:x=2.2,y=-2,z=0)  -- (xyz cs:x=2.3,y=-2,z=1.5)  node[above] {$z$};
			\coordinate (O) at (-1,-1,-3);
			\coordinate (C) at (-0.5,-1,3);
			\coordinate (G) at (-1,1,-3);
			\coordinate (D) at (-0.5,1,3);
			\coordinate (A) at (1,-1,-3);
			\coordinate (B) at (1.5,1,3);
			\coordinate (E) at (1.5,-1,3);
			\coordinate (F) at (1,1,-3);
			\draw[line width=1pt] (O) -- (C) -- (D) -- (G) -- cycle;
			\draw[line width=1pt]  (O) -- (A) -- (E) -- (C) -- cycle;
			\draw[line width=1pt]  (G) -- (D) -- (B) -- (F) -- cycle;
			\draw[line width=1pt]  (F) -- (B) -- (E) -- (A) -- cycle;
		\end{tikzpicture}
	\end{subfigure}
        \begin{subfigure}[b]{0.49\textwidth}
		\centering
		\begin{tikzpicture}
			\coordinate (H) at (1,1);
			\coordinate (I) at (1,-1);
			\coordinate (J) at (-1,1);
			\coordinate (K) at (-1,-1);
			\draw[line width=1pt] (H) -- (J) -- (K) -- (I) -- cycle;
			\draw [<->,line width=1pt] (0,2) node (yaxis) [left]{$y$}  |- (2,0) node (xaxis) [below] {$x$};
			\draw[dashed, <->,line width=1pt] (-1, -1.3) -- (1, -1.3) node[pos=0.5,below]{$2a$};
			\draw[dashed, line width=1pt]   (0,0) -- (1,1) node[pos=0.6,below] {$D$};
		\end{tikzpicture}
        \end{subfigure}
\caption{Problem description: fluid is assumed to be inside a Cuboid which is infinitely long in z direction. At the start of simulation, temperature at walls is suddenly dropped to a value below freezing temperature.}
\label{fig:Cp_H_3}
\end{figure}

The first test case is of solidification in the corner of a square initially filled with fluid, as shown in figure \ref{fig:Cp_H_3}. The cuboid is assumed long enough in the $z$ direction to form a 2D problem at its cross-section. For the solution here, a 3D setup of the domain has been used with a periodic boundary condition in the $z$ direction. Solidification starts as the temperature of all side walls is suddenly changed to a value below the freezing point. From the analytical solution of this problem \cite{Rathjen_71, Budhia_73}, the solid-liquid interface that moves from the corner toward the square centre is of the form

\begin{equation}\label{eq:analytical}
f(x^*)=[\lambda^m+C/(x^{*m}-\lambda^m)]^{1/m}
\end{equation}
where $x^*=x/(4\alpha t)^{1/2}$, and $\lambda$ is calculated by solving

\begin{equation}
\frac{e^{-\lambda^2}}{erf \lambda} - \frac{T^*_i e^{-\lambda^2}}{erfc \lambda} = \sqrt{\pi} \beta \lambda
\end{equation}
where $T_i^*=k_l/k_s  (T_i-T_m)/(T_m-T_w)$ and $\beta=1/ St$. Values of $m$ and $C$ depend on the non-dimensional parameters $T_i^∗$ and $\beta$. For $\beta$=0.25 and $T_i^*$=0.3, these values are $\lambda=0.708$, $C=0.159$ and $m=5.02$ \cite{Rathjen_71}. The non-dimensional position of the solidification front on the cuboid diagonal ($D$ in figure \ref{fig:Cp_H_3}) can be also calculated from equation \ref{eq:analytical}. On the diagonal, $f(x^*)=x^*$. Solving equation \ref{eq:analytical} for this equality results in $f(x^*)=0.8958$.

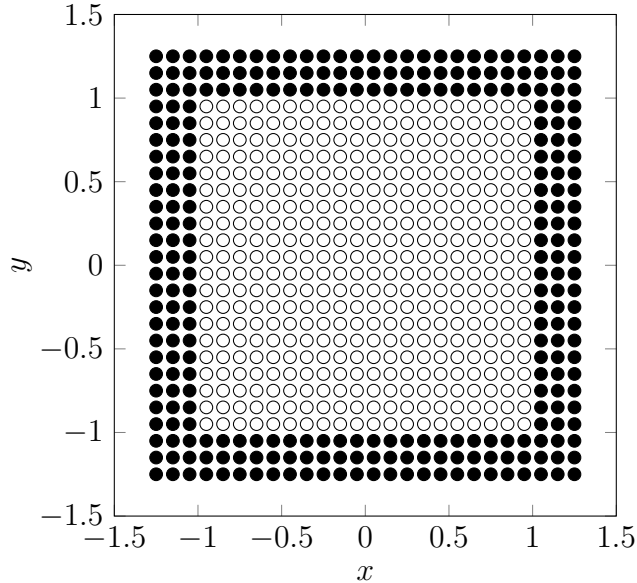
\begin{figure}[h]
        \centering
\begin{tikzpicture}
    \begin{axis}[xlabel=$x$, width=0.5\textwidth,height=0.5\textwidth, ylabel=$y$,cycle list name=black white,tick label style={/pgf/number format/fixed} ]

    \addplot+[smooth,mark=o, mark options={scale=1.2},only marks] 
table[row sep=\\]{
 X Y\\
-9.50E-01	-9.50E-01\\-8.50E-01	-9.50E-01\\-7.50E-01	-9.50E-01\\-6.50E-01	-9.50E-01\\-5.50E-01	-9.50E-01\\-4.50E-01	-9.50E-01\\-3.50E-01	-9.50E-01\\-2.50E-01	-9.50E-01\\-1.50E-01	-9.50E-01\\-5.00E-02	-9.50E-01\\5.00E-02	-9.50E-01\\1.50E-01	-9.50E-01\\2.50E-01	-9.50E-01\\3.50E-01	-9.50E-01\\4.50E-01	-9.50E-01\\5.50E-01	-9.50E-01\\6.50E-01	-9.50E-01\\7.50E-01	-9.50E-01\\8.50E-01	-9.50E-01\\9.50E-01	-9.50E-01\\-9.50E-01	-8.50E-01\\-8.50E-01	-8.50E-01\\-7.50E-01	-8.50E-01\\-6.50E-01	-8.50E-01\\-5.50E-01	-8.50E-01\\-4.50E-01	-8.50E-01\\-3.50E-01	-8.50E-01\\-2.50E-01	-8.50E-01\\-1.50E-01	-8.50E-01\\-5.00E-02	-8.50E-01\\5.00E-02	-8.50E-01\\1.50E-01	-8.50E-01\\2.50E-01	-8.50E-01\\3.50E-01	-8.50E-01\\4.50E-01	-8.50E-01\\5.50E-01	-8.50E-01\\6.50E-01	-8.50E-01\\7.50E-01	-8.50E-01\\8.50E-01	-8.50E-01\\9.50E-01	-8.50E-01\\-9.50E-01	-7.50E-01\\-8.50E-01	-7.50E-01\\-7.50E-01	-7.50E-01\\-6.50E-01	-7.50E-01\\-5.50E-01	-7.50E-01\\-4.50E-01	-7.50E-01\\-3.50E-01	-7.50E-01\\-2.50E-01	-7.50E-01\\-1.50E-01	-7.50E-01\\-5.00E-02	-7.50E-01\\5.00E-02	-7.50E-01\\1.50E-01	-7.50E-01\\2.50E-01	-7.50E-01\\3.50E-01	-7.50E-01\\4.50E-01	-7.50E-01\\5.50E-01	-7.50E-01\\6.50E-01	-7.50E-01\\7.50E-01	-7.50E-01\\8.50E-01	-7.50E-01\\9.50E-01	-7.50E-01\\-9.50E-01	-6.50E-01\\-8.50E-01	-6.50E-01\\-7.50E-01	-6.50E-01\\-6.50E-01	-6.50E-01\\-5.50E-01	-6.50E-01\\-4.50E-01	-6.50E-01\\-3.50E-01	-6.50E-01\\-2.50E-01	-6.50E-01\\-1.50E-01	-6.50E-01\\-5.00E-02	-6.50E-01\\5.00E-02	-6.50E-01\\1.50E-01	-6.50E-01\\2.50E-01	-6.50E-01\\3.50E-01	-6.50E-01\\4.50E-01	-6.50E-01\\5.50E-01	-6.50E-01\\6.50E-01	-6.50E-01\\7.50E-01	-6.50E-01\\8.50E-01	-6.50E-01\\9.50E-01	-6.50E-01\\-9.50E-01	-5.50E-01\\-8.50E-01	-5.50E-01\\-7.50E-01	-5.50E-01\\-6.50E-01	-5.50E-01\\-5.50E-01	-5.50E-01\\-4.50E-01	-5.50E-01\\-3.50E-01	-5.50E-01\\-2.50E-01	-5.50E-01\\-1.50E-01	-5.50E-01\\-5.00E-02	-5.50E-01\\5.00E-02	-5.50E-01\\1.50E-01	-5.50E-01\\2.50E-01	-5.50E-01\\3.50E-01	-5.50E-01\\4.50E-01	-5.50E-01\\5.50E-01	-5.50E-01\\6.50E-01	-5.50E-01\\7.50E-01	-5.50E-01\\8.50E-01	-5.50E-01\\9.50E-01	-5.50E-01\\-9.50E-01	-4.50E-01\\-8.50E-01	-4.50E-01\\-7.50E-01	-4.50E-01\\-6.50E-01	-4.50E-01\\-5.50E-01	-4.50E-01\\-4.50E-01	-4.50E-01\\-3.50E-01	-4.50E-01\\-2.50E-01	-4.50E-01\\-1.50E-01	-4.50E-01\\-5.00E-02	-4.50E-01\\5.00E-02	-4.50E-01\\1.50E-01	-4.50E-01\\2.50E-01	-4.50E-01\\3.50E-01	-4.50E-01\\4.50E-01	-4.50E-01\\5.50E-01	-4.50E-01\\6.50E-01	-4.50E-01\\7.50E-01	-4.50E-01\\8.50E-01	-4.50E-01\\9.50E-01	-4.50E-01\\-9.50E-01	-3.50E-01\\-8.50E-01	-3.50E-01\\-7.50E-01	-3.50E-01\\-6.50E-01	-3.50E-01\\-5.50E-01	-3.50E-01\\-4.50E-01	-3.50E-01\\-3.50E-01	-3.50E-01\\-2.50E-01	-3.50E-01\\-1.50E-01	-3.50E-01\\-5.00E-02	-3.50E-01\\5.00E-02	-3.50E-01\\1.50E-01	-3.50E-01\\2.50E-01	-3.50E-01\\3.50E-01	-3.50E-01\\4.50E-01	-3.50E-01\\5.50E-01	-3.50E-01\\6.50E-01	-3.50E-01\\7.50E-01	-3.50E-01\\8.50E-01	-3.50E-01\\9.50E-01	-3.50E-01\\-9.50E-01	-2.50E-01\\-8.50E-01	-2.50E-01\\-7.50E-01	-2.50E-01\\-6.50E-01	-2.50E-01\\-5.50E-01	-2.50E-01\\-4.50E-01	-2.50E-01\\-3.50E-01	-2.50E-01\\-2.50E-01	-2.50E-01\\-1.50E-01	-2.50E-01\\-5.00E-02	-2.50E-01\\5.00E-02	-2.50E-01\\1.50E-01	-2.50E-01\\2.50E-01	-2.50E-01\\3.50E-01	-2.50E-01\\4.50E-01	-2.50E-01\\5.50E-01	-2.50E-01\\6.50E-01	-2.50E-01\\7.50E-01	-2.50E-01\\8.50E-01	-2.50E-01\\9.50E-01	-2.50E-01\\-9.50E-01	-1.50E-01\\-8.50E-01	-1.50E-01\\-7.50E-01	-1.50E-01\\-6.50E-01	-1.50E-01\\-5.50E-01	-1.50E-01\\-4.50E-01	-1.50E-01\\-3.50E-01	-1.50E-01\\-2.50E-01	-1.50E-01\\-1.50E-01	-1.50E-01\\-5.00E-02	-1.50E-01\\5.00E-02	-1.50E-01\\1.50E-01	-1.50E-01\\2.50E-01	-1.50E-01\\3.50E-01	-1.50E-01\\4.50E-01	-1.50E-01\\5.50E-01	-1.50E-01\\6.50E-01	-1.50E-01\\7.50E-01	-1.50E-01\\8.50E-01	-1.50E-01\\9.50E-01	-1.50E-01\\-9.50E-01	-5.00E-02\\-8.50E-01	-5.00E-02\\-7.50E-01	-5.00E-02\\-6.50E-01	-5.00E-02\\-5.50E-01	-5.00E-02\\-4.50E-01	-5.00E-02\\-3.50E-01	-5.00E-02\\-2.50E-01	-5.00E-02\\-1.50E-01	-5.00E-02\\-5.00E-02	-5.00E-02\\5.00E-02	-5.00E-02\\1.50E-01	-5.00E-02\\2.50E-01	-5.00E-02\\3.50E-01	-5.00E-02\\4.50E-01	-5.00E-02\\5.50E-01	-5.00E-02\\6.50E-01	-5.00E-02\\7.50E-01	-5.00E-02\\8.50E-01	-5.00E-02\\9.50E-01	-5.00E-02\\-9.50E-01	5.00E-02\\-8.50E-01	5.00E-02\\-7.50E-01	5.00E-02\\-6.50E-01	5.00E-02\\-5.50E-01	5.00E-02\\-4.50E-01	5.00E-02\\-3.50E-01	5.00E-02\\-2.50E-01	5.00E-02\\-1.50E-01	5.00E-02\\-5.00E-02	5.00E-02\\5.00E-02	5.00E-02\\1.50E-01	5.00E-02\\2.50E-01	5.00E-02\\3.50E-01	5.00E-02\\4.50E-01	5.00E-02\\5.50E-01	5.00E-02\\6.50E-01	5.00E-02\\7.50E-01	5.00E-02\\8.50E-01	5.00E-02\\9.50E-01	5.00E-02\\-9.50E-01	1.50E-01\\-8.50E-01	1.50E-01\\-7.50E-01	1.50E-01\\-6.50E-01	1.50E-01\\-5.50E-01	1.50E-01\\-4.50E-01	1.50E-01\\-3.50E-01	1.50E-01\\-2.50E-01	1.50E-01\\-1.50E-01	1.50E-01\\-5.00E-02	1.50E-01\\5.00E-02	1.50E-01\\1.50E-01	1.50E-01\\2.50E-01	1.50E-01\\3.50E-01	1.50E-01\\4.50E-01	1.50E-01\\5.50E-01	1.50E-01\\6.50E-01	1.50E-01\\7.50E-01	1.50E-01\\8.50E-01	1.50E-01\\9.50E-01	1.50E-01\\-9.50E-01	2.50E-01\\-8.50E-01	2.50E-01\\-7.50E-01	2.50E-01\\-6.50E-01	2.50E-01\\-5.50E-01	2.50E-01\\-4.50E-01	2.50E-01\\-3.50E-01	2.50E-01\\-2.50E-01	2.50E-01\\-1.50E-01	2.50E-01\\-5.00E-02	2.50E-01\\5.00E-02	2.50E-01\\1.50E-01	2.50E-01\\2.50E-01	2.50E-01\\3.50E-01	2.50E-01\\4.50E-01	2.50E-01\\5.50E-01	2.50E-01\\6.50E-01	2.50E-01\\7.50E-01	2.50E-01\\8.50E-01	2.50E-01\\9.50E-01	2.50E-01\\-9.50E-01	3.50E-01\\-8.50E-01	3.50E-01\\-7.50E-01	3.50E-01\\-6.50E-01	3.50E-01\\-5.50E-01	3.50E-01\\-4.50E-01	3.50E-01\\-3.50E-01	3.50E-01\\-2.50E-01	3.50E-01\\-1.50E-01	3.50E-01\\-5.00E-02	3.50E-01\\5.00E-02	3.50E-01\\1.50E-01	3.50E-01\\2.50E-01	3.50E-01\\3.50E-01	3.50E-01\\4.50E-01	3.50E-01\\5.50E-01	3.50E-01\\6.50E-01	3.50E-01\\7.50E-01	3.50E-01\\8.50E-01	3.50E-01\\9.50E-01	3.50E-01\\-9.50E-01	4.50E-01\\-8.50E-01	4.50E-01\\-7.50E-01	4.50E-01\\-6.50E-01	4.50E-01\\-5.50E-01	4.50E-01\\-4.50E-01	4.50E-01\\-3.50E-01	4.50E-01\\-2.50E-01	4.50E-01\\-1.50E-01	4.50E-01\\-5.00E-02	4.50E-01\\5.00E-02	4.50E-01\\1.50E-01	4.50E-01\\2.50E-01	4.50E-01\\3.50E-01	4.50E-01\\4.50E-01	4.50E-01\\5.50E-01	4.50E-01\\6.50E-01	4.50E-01\\7.50E-01	4.50E-01\\8.50E-01	4.50E-01\\9.50E-01	4.50E-01\\-9.50E-01	5.50E-01\\-8.50E-01	5.50E-01\\-7.50E-01	5.50E-01\\-6.50E-01	5.50E-01\\-5.50E-01	5.50E-01\\-4.50E-01	5.50E-01\\-3.50E-01	5.50E-01\\-2.50E-01	5.50E-01\\-1.50E-01	5.50E-01\\-5.00E-02	5.50E-01\\5.00E-02	5.50E-01\\1.50E-01	5.50E-01\\2.50E-01	5.50E-01\\3.50E-01	5.50E-01\\4.50E-01	5.50E-01\\5.50E-01	5.50E-01\\6.50E-01	5.50E-01\\7.50E-01	5.50E-01\\8.50E-01	5.50E-01\\9.50E-01	5.50E-01\\-9.50E-01	6.50E-01\\-8.50E-01	6.50E-01\\-7.50E-01	6.50E-01\\-6.50E-01	6.50E-01\\-5.50E-01	6.50E-01\\-4.50E-01	6.50E-01\\-3.50E-01	6.50E-01\\-2.50E-01	6.50E-01\\-1.50E-01	6.50E-01\\-5.00E-02	6.50E-01\\5.00E-02	6.50E-01\\1.50E-01	6.50E-01\\2.50E-01	6.50E-01\\3.50E-01	6.50E-01\\4.50E-01	6.50E-01\\5.50E-01	6.50E-01\\6.50E-01	6.50E-01\\7.50E-01	6.50E-01\\8.50E-01	6.50E-01\\9.50E-01	6.50E-01\\-9.50E-01	7.50E-01\\-8.50E-01	7.50E-01\\-7.50E-01	7.50E-01\\-6.50E-01	7.50E-01\\-5.50E-01	7.50E-01\\-4.50E-01	7.50E-01\\-3.50E-01	7.50E-01\\-2.50E-01	7.50E-01\\-1.50E-01	7.50E-01\\-5.00E-02	7.50E-01\\5.00E-02	7.50E-01\\1.50E-01	7.50E-01\\2.50E-01	7.50E-01\\3.50E-01	7.50E-01\\4.50E-01	7.50E-01\\5.50E-01	7.50E-01\\6.50E-01	7.50E-01\\7.50E-01	7.50E-01\\8.50E-01	7.50E-01\\9.50E-01	7.50E-01\\-9.50E-01	8.50E-01\\-8.50E-01	8.50E-01\\-7.50E-01	8.50E-01\\-6.50E-01	8.50E-01\\-5.50E-01	8.50E-01\\-4.50E-01	8.50E-01\\-3.50E-01	8.50E-01\\-2.50E-01	8.50E-01\\-1.50E-01	8.50E-01\\-5.00E-02	8.50E-01\\5.00E-02	8.50E-01\\1.50E-01	8.50E-01\\2.50E-01	8.50E-01\\3.50E-01	8.50E-01\\4.50E-01	8.50E-01\\5.50E-01	8.50E-01\\6.50E-01	8.50E-01\\7.50E-01	8.50E-01\\8.50E-01	8.50E-01\\9.50E-01	8.50E-01\\-9.50E-01	9.50E-01\\-8.50E-01	9.50E-01\\-7.50E-01	9.50E-01\\-6.50E-01	9.50E-01\\-5.50E-01	9.50E-01\\-4.50E-01	9.50E-01\\-3.50E-01	9.50E-01\\-2.50E-01	9.50E-01\\-1.50E-01	9.50E-01\\-5.00E-02	9.50E-01\\5.00E-02	9.50E-01\\1.50E-01	9.50E-01\\2.50E-01	9.50E-01\\3.50E-01	9.50E-01\\4.50E-01	9.50E-01\\5.50E-01	9.50E-01\\6.50E-01	9.50E-01\\7.50E-01	9.50E-01\\8.50E-01	9.50E-01\\9.50E-01	9.50E-01\\
};

    \addplot+[smooth,mark=*, mark options={scale=1.2},only marks] 
table[row sep=\\]{
 X Y\\
-1.25E+00	-1.25E+00\\-1.15E+00	-1.25E+00\\-1.05E+00	-1.25E+00\\-9.50E-01	-1.25E+00\\-8.50E-01	-1.25E+00\\-7.50E-01	-1.25E+00\\-6.50E-01	-1.25E+00\\-5.50E-01	-1.25E+00\\-4.50E-01	-1.25E+00\\-3.50E-01	-1.25E+00\\-2.50E-01	-1.25E+00\\-1.50E-01	-1.25E+00\\-5.00E-02	-1.25E+00\\5.00E-02	-1.25E+00\\1.50E-01	-1.25E+00\\2.50E-01	-1.25E+00\\3.50E-01	-1.25E+00\\4.50E-01	-1.25E+00\\5.50E-01	-1.25E+00\\6.50E-01	-1.25E+00\\7.50E-01	-1.25E+00\\8.50E-01	-1.25E+00\\9.50E-01	-1.25E+00\\1.05E+00	-1.25E+00\\1.15E+00	-1.25E+00\\1.25E+00	-1.25E+00\\-1.25E+00	-1.15E+00\\-1.15E+00	-1.15E+00\\-1.05E+00	-1.15E+00\\-9.50E-01	-1.15E+00\\-8.50E-01	-1.15E+00\\-7.50E-01	-1.15E+00\\-6.50E-01	-1.15E+00\\-5.50E-01	-1.15E+00\\-4.50E-01	-1.15E+00\\-3.50E-01	-1.15E+00\\-2.50E-01	-1.15E+00\\-1.50E-01	-1.15E+00\\-5.00E-02	-1.15E+00\\5.00E-02	-1.15E+00\\1.50E-01	-1.15E+00\\2.50E-01	-1.15E+00\\3.50E-01	-1.15E+00\\4.50E-01	-1.15E+00\\5.50E-01	-1.15E+00\\6.50E-01	-1.15E+00\\7.50E-01	-1.15E+00\\8.50E-01	-1.15E+00\\9.50E-01	-1.15E+00\\1.05E+00	-1.15E+00\\1.15E+00	-1.15E+00\\1.25E+00	-1.15E+00\\-1.25E+00	-1.05E+00\\-1.15E+00	-1.05E+00\\-1.05E+00	-1.05E+00\\-9.50E-01	-1.05E+00\\-8.50E-01	-1.05E+00\\-7.50E-01	-1.05E+00\\-6.50E-01	-1.05E+00\\-5.50E-01	-1.05E+00\\-4.50E-01	-1.05E+00\\-3.50E-01	-1.05E+00\\-2.50E-01	-1.05E+00\\-1.50E-01	-1.05E+00\\-5.00E-02	-1.05E+00\\5.00E-02	-1.05E+00\\1.50E-01	-1.05E+00\\2.50E-01	-1.05E+00\\3.50E-01	-1.05E+00\\4.50E-01	-1.05E+00\\5.50E-01	-1.05E+00\\6.50E-01	-1.05E+00\\7.50E-01	-1.05E+00\\8.50E-01	-1.05E+00\\9.50E-01	-1.05E+00\\1.05E+00	-1.05E+00\\1.15E+00	-1.05E+00\\1.25E+00	-1.05E+00\\-1.25E+00	-9.50E-01\\-1.15E+00	-9.50E-01\\-1.05E+00	-9.50E-01\\1.05E+00	-9.50E-01\\1.15E+00	-9.50E-01\\1.25E+00	-9.50E-01\\-1.25E+00	-8.50E-01\\-1.15E+00	-8.50E-01\\-1.05E+00	-8.50E-01\\1.05E+00	-8.50E-01\\1.15E+00	-8.50E-01\\1.25E+00	-8.50E-01\\-1.25E+00	-7.50E-01\\-1.15E+00	-7.50E-01\\-1.05E+00	-7.50E-01\\1.05E+00	-7.50E-01\\1.15E+00	-7.50E-01\\1.25E+00	-7.50E-01\\-1.25E+00	-6.50E-01\\-1.15E+00	-6.50E-01\\-1.05E+00	-6.50E-01\\1.05E+00	-6.50E-01\\1.15E+00	-6.50E-01\\1.25E+00	-6.50E-01\\-1.25E+00	-5.50E-01\\-1.15E+00	-5.50E-01\\-1.05E+00	-5.50E-01\\1.05E+00	-5.50E-01\\1.15E+00	-5.50E-01\\1.25E+00	-5.50E-01\\-1.25E+00	-4.50E-01\\-1.15E+00	-4.50E-01\\-1.05E+00	-4.50E-01\\1.05E+00	-4.50E-01\\1.15E+00	-4.50E-01\\1.25E+00	-4.50E-01\\-1.25E+00	-3.50E-01\\-1.15E+00	-3.50E-01\\-1.05E+00	-3.50E-01\\1.05E+00	-3.50E-01\\1.15E+00	-3.50E-01\\1.25E+00	-3.50E-01\\-1.25E+00	-2.50E-01\\-1.15E+00	-2.50E-01\\-1.05E+00	-2.50E-01\\1.05E+00	-2.50E-01\\1.15E+00	-2.50E-01\\1.25E+00	-2.50E-01\\-1.25E+00	-1.50E-01\\-1.15E+00	-1.50E-01\\-1.05E+00	-1.50E-01\\1.05E+00	-1.50E-01\\1.15E+00	-1.50E-01\\1.25E+00	-1.50E-01\\-1.25E+00	-5.00E-02\\-1.15E+00	-5.00E-02\\-1.05E+00	-5.00E-02\\1.05E+00	-5.00E-02\\1.15E+00	-5.00E-02\\1.25E+00	-5.00E-02\\-1.25E+00	5.00E-02\\-1.15E+00	5.00E-02\\-1.05E+00	5.00E-02\\1.05E+00	5.00E-02\\1.15E+00	5.00E-02\\1.25E+00	5.00E-02\\-1.25E+00	1.50E-01\\-1.15E+00	1.50E-01\\-1.05E+00	1.50E-01\\1.05E+00	1.50E-01\\1.15E+00	1.50E-01\\1.25E+00	1.50E-01\\-1.25E+00	2.50E-01\\-1.15E+00	2.50E-01\\-1.05E+00	2.50E-01\\1.05E+00	2.50E-01\\1.15E+00	2.50E-01\\1.25E+00	2.50E-01\\-1.25E+00	3.50E-01\\-1.15E+00	3.50E-01\\-1.05E+00	3.50E-01\\1.05E+00	3.50E-01\\1.15E+00	3.50E-01\\1.25E+00	3.50E-01\\-1.25E+00	4.50E-01\\-1.15E+00	4.50E-01\\-1.05E+00	4.50E-01\\1.05E+00	4.50E-01\\1.15E+00	4.50E-01\\1.25E+00	4.50E-01\\-1.25E+00	5.50E-01\\-1.15E+00	5.50E-01\\-1.05E+00	5.50E-01\\1.05E+00	5.50E-01\\1.15E+00	5.50E-01\\1.25E+00	5.50E-01\\-1.25E+00	6.50E-01\\-1.15E+00	6.50E-01\\-1.05E+00	6.50E-01\\1.05E+00	6.50E-01\\1.15E+00	6.50E-01\\1.25E+00	6.50E-01\\-1.25E+00	7.50E-01\\-1.15E+00	7.50E-01\\-1.05E+00	7.50E-01\\1.05E+00	7.50E-01\\1.15E+00	7.50E-01\\1.25E+00	7.50E-01\\-1.25E+00	8.50E-01\\-1.15E+00	8.50E-01\\-1.05E+00	8.50E-01\\1.05E+00	8.50E-01\\1.15E+00	8.50E-01\\1.25E+00	8.50E-01\\-1.25E+00	9.50E-01\\-1.15E+00	9.50E-01\\-1.05E+00	9.50E-01\\1.05E+00	9.50E-01\\1.15E+00	9.50E-01\\1.25E+00	9.50E-01\\-1.25E+00	1.05E+00\\-1.15E+00	1.05E+00\\-1.05E+00	1.05E+00\\-9.50E-01	1.05E+00\\-8.50E-01	1.05E+00\\-7.50E-01	1.05E+00\\-6.50E-01	1.05E+00\\-5.50E-01	1.05E+00\\-4.50E-01	1.05E+00\\-3.50E-01	1.05E+00\\-2.50E-01	1.05E+00\\-1.50E-01	1.05E+00\\-5.00E-02	1.05E+00\\5.00E-02	1.05E+00\\1.50E-01	1.05E+00\\2.50E-01	1.05E+00\\3.50E-01	1.05E+00\\4.50E-01	1.05E+00\\5.50E-01	1.05E+00\\6.50E-01	1.05E+00\\7.50E-01	1.05E+00\\8.50E-01	1.05E+00\\9.50E-01	1.05E+00\\1.05E+00	1.05E+00\\1.15E+00	1.05E+00\\1.25E+00	1.05E+00\\-1.25E+00	1.15E+00\\-1.15E+00	1.15E+00\\-1.05E+00	1.15E+00\\-9.50E-01	1.15E+00\\-8.50E-01	1.15E+00\\-7.50E-01	1.15E+00\\-6.50E-01	1.15E+00\\-5.50E-01	1.15E+00\\-4.50E-01	1.15E+00\\-3.50E-01	1.15E+00\\-2.50E-01	1.15E+00\\-1.50E-01	1.15E+00\\-5.00E-02	1.15E+00\\5.00E-02	1.15E+00\\1.50E-01	1.15E+00\\2.50E-01	1.15E+00\\3.50E-01	1.15E+00\\4.50E-01	1.15E+00\\5.50E-01	1.15E+00\\6.50E-01	1.15E+00\\7.50E-01	1.15E+00\\8.50E-01	1.15E+00\\9.50E-01	1.15E+00\\1.05E+00	1.15E+00\\1.15E+00	1.15E+00\\1.25E+00	1.15E+00\\-1.25E+00	1.25E+00\\-1.15E+00	1.25E+00\\-1.05E+00	1.25E+00\\-9.50E-01	1.25E+00\\-8.50E-01	1.25E+00\\-7.50E-01	1.25E+00\\-6.50E-01	1.25E+00\\-5.50E-01	1.25E+00\\-4.50E-01	1.25E+00\\-3.50E-01	1.25E+00\\-2.50E-01	1.25E+00\\-1.50E-01	1.25E+00\\-5.00E-02	1.25E+00\\5.00E-02	1.25E+00\\1.50E-01	1.25E+00\\2.50E-01	1.25E+00\\3.50E-01	1.25E+00\\4.50E-01	1.25E+00\\5.50E-01	1.25E+00\\6.50E-01	1.25E+00\\7.50E-01	1.25E+00\\8.50E-01	1.25E+00\\9.50E-01	1.25E+00\\1.05E+00	1.25E+00\\1.15E+00	1.25E+00\\1.25E+00	1.25E+00\\
};
    \end{axis}
\end{tikzpicture}
\caption{Initial particle positioning. Hollow circles show fluid particle positions, while filled circles indicate wall particles.}
\label{fig:Cp_4}
\end{figure}

Numerically, SPH particles are uniformly distributed in all directions. Figure \ref{fig:Cp_4} shows the particle distribution at any $x-y$ cross-section. Since each particle has a neighbourhood radius of $3\Delta x$, three layers of particles are placed in the wall so that all inner particles have a complete neighbourhood (filled circles in figure \ref{fig:Cp_4}). The temperature of all wall particles is kept constant throughout the simulation. Properties of the solid and liquid are assumed to remain constant in each phase. Density is assumed to be the same in the two phases.

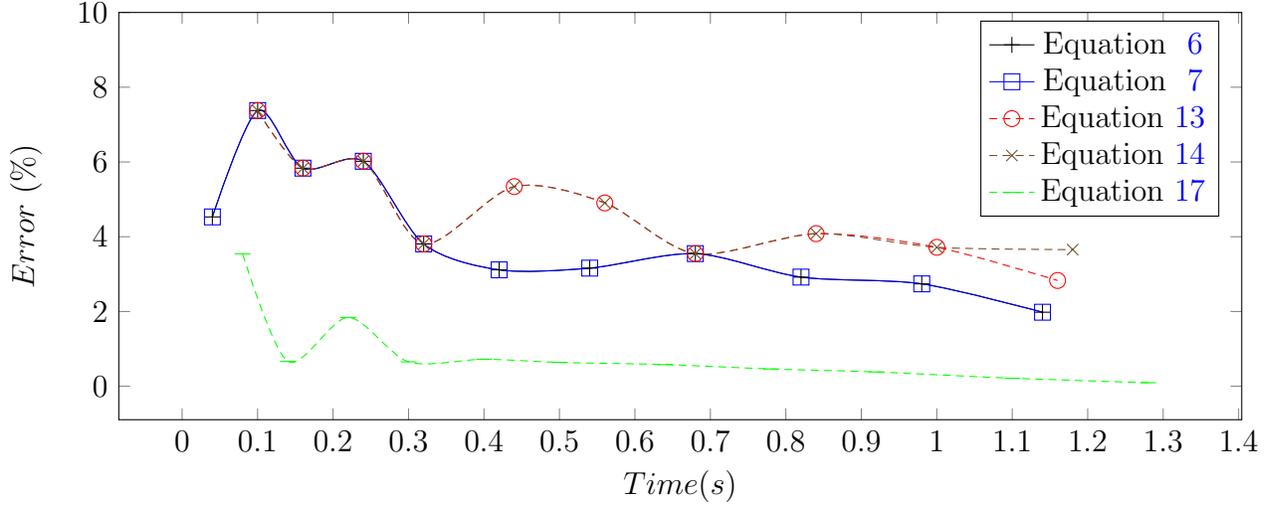
\begin{figure}[h]
        \centering
\begin{tikzpicture}
    \begin{axis}[xlabel=$Time (s)$, width=\textwidth,height=7cm, ylabel=$Error~(\%)$, ymax=10]
\pgfplotsset{cycle list shift=3}

    \addplot+[smooth,mark=+, mark options={scale=1.5, solid}] 
table[row sep=\\]{
 X Y\\
0.0401	4.530317 \\  0.1001	7.375591 \\  0.1601	5.834613 \\  0.2401	6.018785 \\  0.3201	3.806668 \\  0.4201	3.11437 \\  0.5401	3.159665 \\  0.6801	3.547905 \\  0.8201	2.919872 \\  0.9801	2.739367 \\  1.1401	1.980386 \\  
};

    \addplot+[smooth,mark=square, mark options={scale=1.5, solid}] 
table[row sep=\\]{
 X Y\\
0.0401	4.530317 \\  0.1001	7.375591 \\  0.1601	5.834613 \\  0.2401	6.018785 \\  0.3201	3.806668 \\  0.4201	3.11437 \\  0.5401	3.159665 \\  0.6801	3.547905 \\  0.8201	2.919872 \\  0.9801	2.739367 \\  1.1401	1.980386 \\  
};

    \addplot+[smooth,mark=o, mark options={scale=1.5, solid}] 
table[row sep=\\]{
 X Y\\
0.1001	7.375591 \\  0.1601	5.834613 \\  0.2401	6.018785 \\  0.3201	3.806668 \\  0.4401	5.341411 \\  0.5601	4.904364 \\  0.6801	3.547905 \\  0.8401	4.082411 \\  1.0001	3.716787 \\  1.1601	2.828983 \\  
};

    \addplot+[smooth,mark=x, mark options={scale=1.5, solid}] 
table[row sep=\\]{
 X Y\\
0.1001	7.375591 \\  0.1601	5.834613 \\  0.2401	6.018785 \\  0.3201	3.806668 \\  0.4401	5.341411 \\  0.5601	4.904364 \\  0.6801	3.547905 \\  0.8401	4.082411 \\  1.0001	3.716787 \\  1.1801	3.655915 \\  
};

    \addplot+[smooth,mark=-, color=green,mark options={scale=1.5, solid}] 
table[row sep=\\]{
 X Y\\
0.0801	3.544309 \\  0.1401	0.662542 \\  0.2201	1.84168 \\  0.3001	0.652989 \\  0.4001	0.722362 \\  0.5001	0.638683 \\  0.6401	0.579913 \\  0.7801	0.46207 \\  0.9201	0.381752 \\  1.1001	0.214284 \\  1.2801	0.09522 \\  
};

\legend{Equation ~\ref{eq:Cp_20}, Equation ~\ref{eq:Cp_21},Equation \ref{eq:Cp_12}, Equation \ref{eq:Cp_15}, Equation \ref{eq:Cp_18}}
    \end{axis}
\end{tikzpicture}
\caption{Solidification front position versus the analytical solution, for solidification in a corner.}
\label{fig:Cp_7}
\end{figure}

A comparison of numerical tests using each of the five equations versus the analytical solution is shown in figure \ref{fig:Cp_7}. The error values are calculated by comparing the numerical solidification front position on the diagonal against the analytical value of $0.8958$ using $\left|{\frac{\bar{x}^*-0.8958}{0.8958} \times 100}\right|$.

Results demonstrate that equations \ref{eq:Cp_20}, \ref{eq:Cp_21}, \ref{eq:Cp_12}, and \ref{eq:Cp_15} yield nearly identical large errors, while the most accurate results are obtained using equation \ref{eq:Cp_18}.

\subsubsection{Comparison to Numerical Results}

\begin{figure}[h]
        \centering
\begin{tikzpicture}[spy using outlines={black,magnification=4,size=3cm, connect spies}]
    \begin{axis}[xlabel=$t\alpha/a^2$, width=\textwidth,height=7cm, ylabel=$l/D$,xmin=0,xmax=0.45,ymin=0,tick label style={/pgf/number format/fixed} ]

\spy on (4.5,3) in node at (7.75,1.75);

    \addplot+[smooth,mark=none,dashed] 
table[row sep=\\]{
 X Y\\
0	0.00233246 \\  7.81E-07	0.00234374 \\  0.00261171	0.0400543 \\  0.00381679	0.0574596 \\  0.00441991	0.0661707 \\  0.00462303	0.0691045 \\  0.00642694	0.0917204 \\  0.00722889	0.101775 \\  0.00863982	0.119464 \\  0.0088367	0.12176 \\  0.0104414	0.14048 \\  0.0112445	0.149848 \\  0.0126554	0.166307 \\  0.0136547	0.176511 \\  0.0154582	0.194927 \\  0.0158578	0.199007 \\  0.0182746	0.223685 \\  0.0190679	0.230422 \\  0.0196699	0.235533 \\  0.0204808	0.242418 \\  0.0226777	0.258452 \\  0.0236792	0.265762 \\  0.0236886	0.265831 \\  0.0260839	0.277489 \\  0.0272945	0.283381 \\  0.0280874	0.288012 \\  0.0312941	0.306739 \\  0.0320948	0.311415 \\  0.0341077	0.323171 \\  0.0360999	0.330806 \\  0.0393069	0.343097 \\  0.0405159	0.34773 \\  0.0419097	0.35293 \\  0.0419128	0.352942 \\  0.0431124	0.357418 \\  0.0459147	0.367873 \\  0.0477249	0.374627 \\  0.048521	0.377121 \\  0.0509206	0.384639 \\  0.0517229	0.387153 \\  0.0559334	0.400344 \\  0.0563264	0.401491 \\  0.0577303	0.405589 \\  0.0621327	0.41844 \\  0.0623397	0.419044 \\  0.0625377	0.419678 \\  0.0635366	0.422877 \\  0.0685463	0.438918 \\  0.0687389	0.439431 \\  0.0735467	0.45226 \\  0.0755525	0.457613 \\  0.0759518	0.458778 \\  0.0785482	0.466357 \\  0.0803572	0.471638 \\  0.0809529	0.47306 \\  0.0843584	0.48119 \\  0.0871576	0.487872 \\  0.0891642	0.492663 \\  0.091758	0.499788 \\  0.093963	0.505845 \\  0.0949665	0.508602 \\  0.0959705	0.51136 \\  0.0993638	0.521548 \\  0.101968	0.529369 \\  0.102977	0.532399 \\  0.103974	0.535154 \\  0.10617	0.541228 \\  0.106781	0.542916 \\  0.107574	0.545093 \\  0.112181	0.55775 \\  0.113587	0.561613 \\  0.114778	0.564592 \\  0.117791	0.572127 \\  0.11818	0.572953 \\  0.12299	0.583156 \\  0.124186	0.585693 \\  0.126597	0.590808 \\  0.127988	0.592512 \\  0.130398	0.595466 \\  0.131391	0.597167 \\  0.132197	0.598548 \\  0.137202	0.607131 \\  0.138195	0.608826 \\  0.138801	0.609863 \\  0.140397	0.612589 \\  0.146801	0.623531 \\  0.148809	0.626962 \\  0.151811	0.632109 \\  0.152204	0.632783 \\  0.159015	0.644461 \\  0.159409	0.645335 \\  0.160616	0.648012 \\  0.167814	0.66397 \\  0.169022	0.666648 \\  0.171216	0.670986 \\  0.172223	0.672977 \\  0.17662	0.68167 \\  0.18043	0.689203 \\  0.18083	0.689994 \\  0.190235	0.706326 \\  0.190836	0.707607 \\  0.196833	0.720391 \\  0.200643	0.728511 \\  0.201044	0.729364 \\  0.204839	0.737441 \\  0.210442	0.749368 \\  0.215453	0.760036 \\  0.216847	0.763003 \\  0.223651	0.777495 \\  0.228054	0.786872 \\  0.228062	0.78689 \\  0.235058	0.801124 \\  0.236459	0.803975 \\  0.236668	0.804401 \\  0.245664	0.823995 \\  0.250868	0.835331 \\  0.251679	0.837096 \\  
    };

    \addplot+[smooth,color=green,mark=none,dashdotted] 
table[row sep=\\]{
 X Y\\
0	2.22E-16 \\  7.81E-07	1.06E-05 \\  0.00261171	0.0352949 \\  0.00381679	0.0515804 \\  0.00441991	0.0597311 \\  0.00462303	0.0620469 \\  0.00642694	0.0826134 \\  0.00722889	0.0917565 \\  0.00863982	0.107843 \\  0.0088367	0.110087 \\  0.0104414	0.128029 \\  0.0112445	0.137008 \\  0.0126554	0.152784 \\  0.0136547	0.163956 \\  0.0154582	0.177645 \\  0.0158578	0.180678 \\  0.0182746	0.199021 \\  0.0190679	0.205043 \\  0.0196699	0.209612 \\  0.0204808	0.214939 \\  0.0226777	0.229372 \\  0.0236792	0.235952 \\  0.0236886	0.236013 \\  0.0260839	0.251749 \\  0.0272945	0.258348 \\  0.0280874	0.26267 \\  0.0312941	0.28015 \\  0.0320948	0.284515 \\  0.0341077	0.294638 \\  0.0360999	0.304658 \\  0.0393069	0.320787 \\  0.0405159	0.325989 \\  0.0419097	0.331987 \\  0.0419128	0.332001 \\  0.0431124	0.337163 \\  0.0459147	0.347439 \\  0.0477249	0.354076 \\  0.048521	0.356996 \\  0.0509206	0.365794 \\  0.0517229	0.368737 \\  0.0559334	0.384309 \\  0.0563264	0.385763 \\  0.0577303	0.390955 \\  0.0621327	0.405137 \\  0.0623397	0.405804 \\  0.0625377	0.406442 \\  0.0635366	0.40966 \\  0.0685463	0.424869 \\  0.0687389	0.425454 \\  0.0735467	0.44005 \\  0.0755525	0.445113 \\  0.0759518	0.44612 \\  0.0785482	0.452674 \\  0.0803572	0.457239 \\  0.0809529	0.458743 \\  0.0843584	0.466434 \\  0.0871576	0.472757 \\  0.0891642	0.477578 \\  0.091758	0.483811 \\  0.093963	0.48911 \\  0.0949665	0.491158 \\  0.0959705	0.493208 \\  0.0993638	0.500136 \\  0.101968	0.505453 \\  0.102977	0.508188 \\  0.103974	0.510888 \\  0.10617	0.516841 \\  0.106781	0.518496 \\  0.107574	0.520644 \\  0.112181	0.529601 \\  0.113587	0.532335 \\  0.114778	0.53465 \\  0.117791	0.541757 \\  0.11818	0.542675 \\  0.12299	0.554022 \\  0.124186	0.556842 \\  0.126597	0.561922 \\  0.127988	0.564852 \\  0.130398	0.56993 \\  0.131391	0.57202 \\  0.132197	0.573587 \\  0.137202	0.583319 \\  0.138195	0.585247 \\  0.138801	0.586427 \\  0.140397	0.589528 \\  0.146801	0.601555 \\  0.148809	0.605326 \\  0.151811	0.610964 \\  0.152204	0.611702 \\  0.159015	0.624942 \\  0.159409	0.625709 \\  0.160616	0.628056 \\  0.167814	0.642048 \\  0.169022	0.644451 \\  0.171216	0.648813 \\  0.172223	0.650816 \\  0.17662	0.659558 \\  0.18043	0.667545 \\  0.18083	0.668384 \\  0.190235	0.687938 \\  0.190836	0.689075 \\  0.196833	0.700431 \\  0.200643	0.707643 \\  0.201044	0.708402 \\  0.204839	0.715587 \\  0.210442	0.727026 \\  0.215453	0.737258 \\  0.216847	0.740103 \\  0.223651	0.752857 \\  0.228054	0.761109 \\  0.228062	0.761124 \\  0.235058	0.773749 \\  0.236459	0.776278 \\  0.236668	0.776702 \\  0.245664	0.794921 \\  0.250868	0.805462 \\  0.251679	0.807104 \\  
    };

    \addplot+[smooth,mark=none] 
table[row sep=\\]{
 X Y\\
0	2.22E-16 \\7.81E-07	1.03E-05 \\0.00261171	0.0344604 \\0.00381679	0.0503608 \\0.00441991	0.0567775 \\0.00462303	0.0589385 \\0.00642694	0.0781303 \\0.00722889	0.0866623 \\0.00863982	0.103121 \\0.0088367	0.105418 \\0.0104414	0.124137 \\0.0112445	0.133506 \\0.0126554	0.147229 \\0.0136547	0.156947 \\0.0154582	0.174489 \\0.0158578	0.177561 \\0.0182746	0.196146 \\0.0190679	0.202246 \\0.0196699	0.206875 \\0.0204808	0.213111 \\0.0226777	0.230004 \\0.0236792	0.237706 \\0.0236886	0.23776 \\0.0260839	0.251749 \\0.0272945	0.258819 \\0.0280874	0.26345 \\0.0312941	0.282177 \\0.0320948	0.286413 \\0.0341077	0.29706 \\0.0360999	0.307598 \\0.0393069	0.324561 \\0.0405159	0.330956 \\0.0419097	0.338328 \\0.0419128	0.338344 \\0.0431124	0.342802 \\0.0459147	0.353217 \\0.0477249	0.359944 \\0.048521	0.362902 \\0.0509206	0.37151 \\0.0517229	0.374388 \\0.0559334	0.389492 \\0.0563264	0.390902 \\0.0577303	0.395938 \\0.0621327	0.41173 \\0.0623397	0.412473 \\0.0625377	0.413183 \\0.0635366	0.416317 \\0.0685463	0.432034 \\0.0687389	0.432638 \\0.0735467	0.447721 \\0.0755525	0.454014 \\0.0759518	0.455266 \\0.0785482	0.461399 \\0.0803572	0.465672 \\0.0809529	0.467079 \\0.0843584	0.475122 \\0.0871576	0.481594 \\0.0891642	0.486233 \\0.091758	0.49223 \\0.093963	0.497329 \\0.0949665	0.499649 \\0.0959705	0.502122 \\0.0993638	0.510484 \\0.101968	0.516902 \\0.102977	0.519389 \\0.103974	0.521844 \\0.10617	0.527472 \\0.106781	0.529037 \\0.107574	0.531068 \\0.112181	0.542874 \\0.113587	0.546369 \\0.114778	0.54933 \\0.117791	0.556818 \\0.11818	0.557786 \\0.12299	0.569742 \\0.124186	0.57232 \\0.126597	0.577519 \\0.127988	0.580518 \\0.130398	0.585716 \\0.131391	0.587855 \\0.132197	0.589592 \\0.137202	0.600211 \\0.138195	0.602316 \\0.138801	0.603603 \\0.140397	0.607183 \\0.146801	0.621555 \\0.148809	0.626062 \\0.151811	0.632798 \\0.152204	0.633527 \\0.159015	0.646164 \\0.159409	0.646896 \\0.160616	0.649136 \\0.167814	0.66216 \\0.169022	0.664346 \\0.171216	0.668316 \\0.172223	0.670139 \\0.17662	0.680777 \\0.18043	0.689997 \\0.18083	0.690714 \\0.190235	0.707588 \\0.190836	0.708665 \\0.196833	0.7217 \\0.200643	0.72998 \\0.201044	0.730851 \\0.204839	0.7391 \\0.210442	0.751278 \\0.215453	0.762171 \\0.216847	0.765199 \\0.223651	0.779989 \\0.228054	0.789558 \\0.228062	0.789577 \\0.235058	0.804781 \\0.236459	0.807826 \\0.236668	0.808281 \\0.245664	0.827834 \\0.250868	0.839146 \\0.251679	0.840908 \\
    };
    
    \addplot+[smooth,mark=+, mark options={scale=1.5, solid},only marks] 
table[row sep=\\]{
9E-05	0.01 \\  0.00549	0.11 \\  0.01089	0.15 \\  0.01629	0.19 \\  0.02169	0.21 \\  0.02709	0.25 \\  0.03249	0.27 \\  0.03789	0.29 \\  0.04329	0.31 \\  0.04869	0.33 \\  0.05409	0.35 \\  0.05949	0.37 \\  0.06489	0.39 \\  0.07029	0.41 \\  0.07569	0.43 \\  	 \\  0.08649	0.45 \\  0.09189	0.47 \\  0.09729	0.49 \\  	 \\  0.10809	0.51 \\  0.11349	0.53 \\  	 \\  0.12429	0.55 \\  	 \\  0.13509	0.57 \\  0.14049	0.59 \\  	 \\  0.15129	0.61 \\  	 \\  0.16209	0.63 \\  	 \\  0.17289	0.65 \\  0.17829	0.67 \\  	 \\  0.18909	0.69 \\  	 \\  0.19989	0.71 \\  	 \\  0.21069	0.73 \\  	 \\  0.22149	0.75 \\  	 \\  0.23229	0.77 \\  0.23769	0.79 \\  	 \\  0.24849	0.81 \\  	 \\  0.25929	0.83 \\  	 \\  0.27009	0.85 \\  0.27549	0.87 \\  	 \\  0.28629	0.89 \\  	 \\  0.29709	0.91 \\  0.30249	0.93 \\  0.30789	0.95 \\  0.31329	0.97 \\  
};

    \addplot+[smooth,mark=square, mark options={scale=1.5, solid},only marks] 
table[row sep=\\]{
9E-05	0.01 \\  0.00549	0.11 \\  0.01089	0.15 \\  0.01629	0.19 \\  0.02169	0.21 \\  0.02709	0.25 \\  0.03249	0.27 \\  0.03789	0.29 \\  0.04329	0.31 \\  0.04869	0.33 \\  0.05409	0.35 \\  0.05949	0.37 \\  0.06489	0.39 \\  0.07029	0.41 \\  0.07569	0.43 \\  	 \\  0.08649	0.45 \\  0.09189	0.47 \\  0.09729	0.49 \\  	 \\  0.10809	0.51 \\  0.11349	0.53 \\  	 \\  0.12429	0.55 \\  	 \\  0.13509	0.57 \\  0.14049	0.59 \\  	 \\  0.15129	0.61 \\  	 \\  0.16209	0.63 \\  	 \\  0.17289	0.65 \\  0.17829	0.67 \\  	 \\  0.18909	0.69 \\  	 \\  0.19989	0.71 \\  	 \\  0.21069	0.73 \\  	 \\  0.22149	0.75 \\  	 \\  0.23229	0.77 \\  0.23769	0.79 \\  	 \\  0.24849	0.81 \\  	 \\  0.25929	0.83 \\  	 \\  0.27009	0.85 \\  0.27549	0.87 \\  	 \\  0.28629	0.89 \\  	 \\  0.29709	0.91 \\  0.30249	0.93 \\  0.30789	0.95 \\  0.31329	0.97 \\  
};

    \addplot+[smooth,mark=o, mark options={scale=1.5, solid},only marks] 
table[row sep=\\]{
 X Y\\
9E-05	0.01 \\  0.00549	0.09 \\  0.01089	0.13 \\  0.01629	0.17 \\  0.02169	0.21 \\  0.02709	0.23 \\  0.03249	0.27 \\  0.03789	0.29 \\  0.04329	0.31 \\  0.04869	0.33 \\  0.05409	0.35 \\  0.05949	0.37 \\    0.07029	0.39 \\  0.07569	0.41 \\  0.08109	0.43 \\  0.08649	0.45  \\  0.09729	0.47 \\  0.10269	0.49  \\  0.11349	0.51 \\  0.11889	0.53 \\    0.12969	0.55 \\  0.13509	0.57 \\   0.14589	0.59 \\    0.15669	0.61 \\   0.16749	0.63 \\  0.17289	0.65 \\  0.18369	0.67 \\   0.19449	0.69 \\    0.20529	0.71 \\  0.21609	0.73 \\    0.22689	0.75 \\  0.23229	0.77 \\    0.24309	0.79 \\    0.25389	0.81 \\    0.26469	0.83 \\   0.27549	0.85 \\   0.28629	0.87 \\  0.29169	0.89 \\    0.30249	0.91 \\  0.30789	0.93 \\    0.31869	0.97 \\
};

    \addplot+[smooth,mark=x, mark options={scale=1.5, solid},only marks] 
table[row sep=\\]{
 X Y\\
9E-05	0.01 \\  0.00549	0.09 \\  0.01089	0.13 \\  0.01629	0.17 \\  0.02169	0.21 \\  0.02709	0.23 \\  0.03249	0.25 \\  0.03789	0.29 \\  0.04329	0.31 \\  0.04869	0.33 \\  0.05409	0.35 \\  0.05949	0.37 \\  	 \\  0.07029	0.39 \\  0.07569	0.41 \\  0.08109	0.43 \\  0.08649	0.45 \\  	 \\  0.09729	0.47 \\  0.10269	0.49 \\  	 \\  0.11349	0.51 \\  0.11889	0.53 \\  	 \\  0.12969	0.55 \\  0.13509	0.57 \\  	 \\  0.14589	0.59 \\  	 \\  0.15669	0.61 \\  	 \\  0.16749	0.63 \\  0.17289	0.65 \\  	 \\  0.18369	0.67 \\  	 \\  0.19449	0.69 \\  	 \\  0.20529	0.71 \\  	 \\  0.21609	0.73 \\  	 \\  0.22689	0.75 \\  0.23229	0.77 \\  	 \\  0.24309	0.79 \\  	 \\  0.25389	0.81 \\  	 \\  0.26469	0.83 \\  	 \\  0.27549	0.85 \\  	 \\  0.28629	0.87 \\  0.29169	0.89 \\  	 \\  0.30249	0.91 \\  0.30789	0.93 \\  	 \\  0.31869	0.95 \\  0.32409	0.97 \\
};

    \addplot+[smooth,mark=-, color=green,mark options={scale=1.5, solid},only marks] 
table[row sep=\\]{
9E-05	0.01 \\  0.00549	0.09 \\  0.01089	0.15 \\  0.01629	0.19 \\  0.02169	0.21 \\  0.02709	0.25 \\  0.03249	0.27 \\  0.03789	0.29 \\  0.04329	0.31 \\  0.04869	0.33 \\  0.05409	0.35 \\  0.05949	0.37 \\  0.06489	0.39 \\  0.07029	0.41 \\  0.07569	0.43 \\  	 \\  0.08649	0.45 \\  0.09189	0.47 \\  0.09729	0.49 \\  	 \\  0.10809	0.51 \\  0.11349	0.53 \\  	 \\  0.12429	0.55 \\  0.12969	0.57 \\  	 \\  0.14049	0.59 \\  	 \\  0.15129	0.61 \\  0.15669	0.63 \\  	 \\  0.16749	0.65 \\  	 \\  0.17829	0.67 \\  	 \\  0.18909	0.69 \\  	 \\  0.19989	0.71 \\  	 \\  0.21069	0.73 \\  0.21609	0.75 \\  	 \\  	 \\  0.23229	0.77 \\  0.23769	0.79 \\  	 \\  0.24849	0.81 \\  	 \\  0.25929	0.83 \\  	 \\  0.27009	0.85 \\  	 \\  0.28089	0.87 \\  0.28629	0.89 \\  	 \\  0.29709	0.91 \\  	 \\  0.30789	0.93 \\  0.31329	0.95 \\  0.31869	0.97 \\  
};

\legend{Hsiao \cite{Hsiao_86}~~~~~, Cao \cite{Cao_89}~~~~~~, Keung \cite{Keung_80}~~, Equation ~\ref{eq:Cp_20},Equation ~\ref{eq:Cp_21}, Equation \ref{eq:Cp_12},Equation \ref{eq:Cp_15},Equation \ref{eq:Cp_18}}
    \end{axis}
\end{tikzpicture}
\caption{Position of the solidification front on the diagonal cross section of the cuboid (resolution: $100 \times 100$)}
\label{fig:Cp_5}
\end{figure}
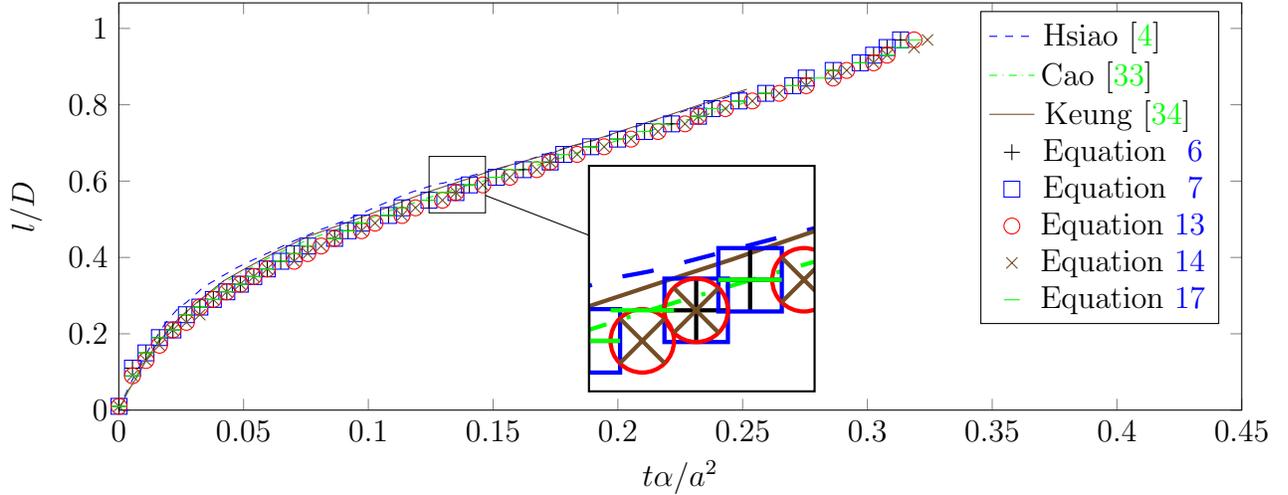

To compare the five equations against available numerical results, the solidification in the corner region is solved for the following non-dimensional parameters: $\theta=(T_i-T_w )/(T_m-T_w )=9/7$, $St=C_s (T_m-T_w )/L=2$, $\alpha_l/\alpha_s=0.9$, and $k_l/k_s=0.9$. The tests conducted here have a fine particle resolution of $100 \times 100$ particles in the $x-y$ cross-section. Results for this case are plotted in figure \ref{fig:Cp_5}. Numerical results of the same problem from Cao et al. \cite{Cao_89}, Hsiao et al. \cite{Hsiao_86}, and Keung \cite{Keung_80} using a coarser mesh of $20 \times 20$ are also shown. All methods converge to the same solution.

\subsubsection{Effect of particle resolution}

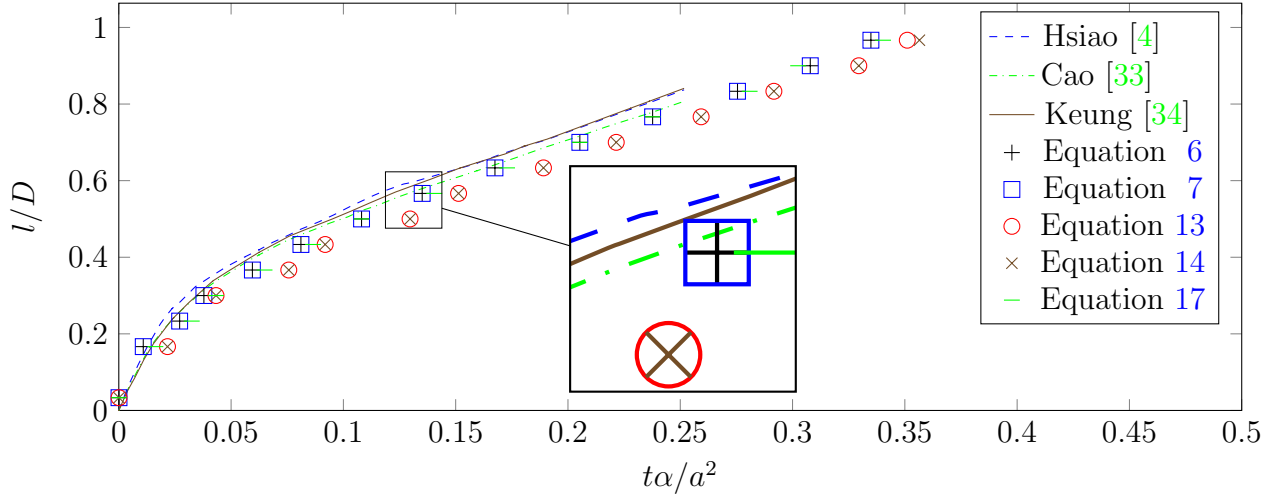
\begin{figure}[h]
        \centering
\begin{tikzpicture}[spy using outlines={black,magnification=4,size=3cm, connect spies}]
    \begin{axis}[xlabel=$t\alpha/a^2$, width=\textwidth,height=7cm, ylabel=$l/D$,xmin=0,xmax=0.5,ymin=0,tick label style={/pgf/number format/fixed}]

\spy on (3.92,2.8) in node at (7.5,1.75);

    \addplot+[smooth,mark=none,dashed] 
table[row sep=\\]{
 X Y\\
0	0.00233246 \\  7.81E-07	0.00234374 \\  0.00261171	0.0400543 \\  0.00381679	0.0574596 \\  0.00441991	0.0661707 \\  0.00462303	0.0691045 \\  0.00642694	0.0917204 \\  0.00722889	0.101775 \\  0.00863982	0.119464 \\  0.0088367	0.12176 \\  0.0104414	0.14048 \\  0.0112445	0.149848 \\  0.0126554	0.166307 \\  0.0136547	0.176511 \\  0.0154582	0.194927 \\  0.0158578	0.199007 \\  0.0182746	0.223685 \\  0.0190679	0.230422 \\  0.0196699	0.235533 \\  0.0204808	0.242418 \\  0.0226777	0.258452 \\  0.0236792	0.265762 \\  0.0236886	0.265831 \\  0.0260839	0.277489 \\  0.0272945	0.283381 \\  0.0280874	0.288012 \\  0.0312941	0.306739 \\  0.0320948	0.311415 \\  0.0341077	0.323171 \\  0.0360999	0.330806 \\  0.0393069	0.343097 \\  0.0405159	0.34773 \\  0.0419097	0.35293 \\  0.0419128	0.352942 \\  0.0431124	0.357418 \\  0.0459147	0.367873 \\  0.0477249	0.374627 \\  0.048521	0.377121 \\  0.0509206	0.384639 \\  0.0517229	0.387153 \\  0.0559334	0.400344 \\  0.0563264	0.401491 \\  0.0577303	0.405589 \\  0.0621327	0.41844 \\  0.0623397	0.419044 \\  0.0625377	0.419678 \\  0.0635366	0.422877 \\  0.0685463	0.438918 \\  0.0687389	0.439431 \\  0.0735467	0.45226 \\  0.0755525	0.457613 \\  0.0759518	0.458778 \\  0.0785482	0.466357 \\  0.0803572	0.471638 \\  0.0809529	0.47306 \\  0.0843584	0.48119 \\  0.0871576	0.487872 \\  0.0891642	0.492663 \\  0.091758	0.499788 \\  0.093963	0.505845 \\  0.0949665	0.508602 \\  0.0959705	0.51136 \\  0.0993638	0.521548 \\  0.101968	0.529369 \\  0.102977	0.532399 \\  0.103974	0.535154 \\  0.10617	0.541228 \\  0.106781	0.542916 \\  0.107574	0.545093 \\  0.112181	0.55775 \\  0.113587	0.561613 \\  0.114778	0.564592 \\  0.117791	0.572127 \\  0.11818	0.572953 \\  0.12299	0.583156 \\  0.124186	0.585693 \\  0.126597	0.590808 \\  0.127988	0.592512 \\  0.130398	0.595466 \\  0.131391	0.597167 \\  0.132197	0.598548 \\  0.137202	0.607131 \\  0.138195	0.608826 \\  0.138801	0.609863 \\  0.140397	0.612589 \\  0.146801	0.623531 \\  0.148809	0.626962 \\  0.151811	0.632109 \\  0.152204	0.632783 \\  0.159015	0.644461 \\  0.159409	0.645335 \\  0.160616	0.648012 \\  0.167814	0.66397 \\  0.169022	0.666648 \\  0.171216	0.670986 \\  0.172223	0.672977 \\  0.17662	0.68167 \\  0.18043	0.689203 \\  0.18083	0.689994 \\  0.190235	0.706326 \\  0.190836	0.707607 \\  0.196833	0.720391 \\  0.200643	0.728511 \\  0.201044	0.729364 \\  0.204839	0.737441 \\  0.210442	0.749368 \\  0.215453	0.760036 \\  0.216847	0.763003 \\  0.223651	0.777495 \\  0.228054	0.786872 \\  0.228062	0.78689 \\  0.235058	0.801124 \\  0.236459	0.803975 \\  0.236668	0.804401 \\  0.245664	0.823995 \\  0.250868	0.835331 \\  0.251679	0.837096 \\  
    };

    \addplot+[smooth,color=green,mark=none,dashdotted] 
table[row sep=\\]{
 X Y\\
0	2.22E-16 \\  7.81E-07	1.06E-05 \\  0.00261171	0.0352949 \\  0.00381679	0.0515804 \\  0.00441991	0.0597311 \\  0.00462303	0.0620469 \\  0.00642694	0.0826134 \\  0.00722889	0.0917565 \\  0.00863982	0.107843 \\  0.0088367	0.110087 \\  0.0104414	0.128029 \\  0.0112445	0.137008 \\  0.0126554	0.152784 \\  0.0136547	0.163956 \\  0.0154582	0.177645 \\  0.0158578	0.180678 \\  0.0182746	0.199021 \\  0.0190679	0.205043 \\  0.0196699	0.209612 \\  0.0204808	0.214939 \\  0.0226777	0.229372 \\  0.0236792	0.235952 \\  0.0236886	0.236013 \\  0.0260839	0.251749 \\  0.0272945	0.258348 \\  0.0280874	0.26267 \\  0.0312941	0.28015 \\  0.0320948	0.284515 \\  0.0341077	0.294638 \\  0.0360999	0.304658 \\  0.0393069	0.320787 \\  0.0405159	0.325989 \\  0.0419097	0.331987 \\  0.0419128	0.332001 \\  0.0431124	0.337163 \\  0.0459147	0.347439 \\  0.0477249	0.354076 \\  0.048521	0.356996 \\  0.0509206	0.365794 \\  0.0517229	0.368737 \\  0.0559334	0.384309 \\  0.0563264	0.385763 \\  0.0577303	0.390955 \\  0.0621327	0.405137 \\  0.0623397	0.405804 \\  0.0625377	0.406442 \\  0.0635366	0.40966 \\  0.0685463	0.424869 \\  0.0687389	0.425454 \\  0.0735467	0.44005 \\  0.0755525	0.445113 \\  0.0759518	0.44612 \\  0.0785482	0.452674 \\  0.0803572	0.457239 \\  0.0809529	0.458743 \\  0.0843584	0.466434 \\  0.0871576	0.472757 \\  0.0891642	0.477578 \\  0.091758	0.483811 \\  0.093963	0.48911 \\  0.0949665	0.491158 \\  0.0959705	0.493208 \\  0.0993638	0.500136 \\  0.101968	0.505453 \\  0.102977	0.508188 \\  0.103974	0.510888 \\  0.10617	0.516841 \\  0.106781	0.518496 \\  0.107574	0.520644 \\  0.112181	0.529601 \\  0.113587	0.532335 \\  0.114778	0.53465 \\  0.117791	0.541757 \\  0.11818	0.542675 \\  0.12299	0.554022 \\  0.124186	0.556842 \\  0.126597	0.561922 \\  0.127988	0.564852 \\  0.130398	0.56993 \\  0.131391	0.57202 \\  0.132197	0.573587 \\  0.137202	0.583319 \\  0.138195	0.585247 \\  0.138801	0.586427 \\  0.140397	0.589528 \\  0.146801	0.601555 \\  0.148809	0.605326 \\  0.151811	0.610964 \\  0.152204	0.611702 \\  0.159015	0.624942 \\  0.159409	0.625709 \\  0.160616	0.628056 \\  0.167814	0.642048 \\  0.169022	0.644451 \\  0.171216	0.648813 \\  0.172223	0.650816 \\  0.17662	0.659558 \\  0.18043	0.667545 \\  0.18083	0.668384 \\  0.190235	0.687938 \\  0.190836	0.689075 \\  0.196833	0.700431 \\  0.200643	0.707643 \\  0.201044	0.708402 \\  0.204839	0.715587 \\  0.210442	0.727026 \\  0.215453	0.737258 \\  0.216847	0.740103 \\  0.223651	0.752857 \\  0.228054	0.761109 \\  0.228062	0.761124 \\  0.235058	0.773749 \\  0.236459	0.776278 \\  0.236668	0.776702 \\  0.245664	0.794921 \\  0.250868	0.805462 \\  0.251679	0.807104 \\  
    };

    \addplot+[smooth,mark=none] 
table[row sep=\\]{
 X Y\\
0	2.22E-16 \\7.81E-07	1.03E-05 \\0.00261171	0.0344604 \\0.00381679	0.0503608 \\0.00441991	0.0567775 \\0.00462303	0.0589385 \\0.00642694	0.0781303 \\0.00722889	0.0866623 \\0.00863982	0.103121 \\0.0088367	0.105418 \\0.0104414	0.124137 \\0.0112445	0.133506 \\0.0126554	0.147229 \\0.0136547	0.156947 \\0.0154582	0.174489 \\0.0158578	0.177561 \\0.0182746	0.196146 \\0.0190679	0.202246 \\0.0196699	0.206875 \\0.0204808	0.213111 \\0.0226777	0.230004 \\0.0236792	0.237706 \\0.0236886	0.23776 \\0.0260839	0.251749 \\0.0272945	0.258819 \\0.0280874	0.26345 \\0.0312941	0.282177 \\0.0320948	0.286413 \\0.0341077	0.29706 \\0.0360999	0.307598 \\0.0393069	0.324561 \\0.0405159	0.330956 \\0.0419097	0.338328 \\0.0419128	0.338344 \\0.0431124	0.342802 \\0.0459147	0.353217 \\0.0477249	0.359944 \\0.048521	0.362902 \\0.0509206	0.37151 \\0.0517229	0.374388 \\0.0559334	0.389492 \\0.0563264	0.390902 \\0.0577303	0.395938 \\0.0621327	0.41173 \\0.0623397	0.412473 \\0.0625377	0.413183 \\0.0635366	0.416317 \\0.0685463	0.432034 \\0.0687389	0.432638 \\0.0735467	0.447721 \\0.0755525	0.454014 \\0.0759518	0.455266 \\0.0785482	0.461399 \\0.0803572	0.465672 \\0.0809529	0.467079 \\0.0843584	0.475122 \\0.0871576	0.481594 \\0.0891642	0.486233 \\0.091758	0.49223 \\0.093963	0.497329 \\0.0949665	0.499649 \\0.0959705	0.502122 \\0.0993638	0.510484 \\0.101968	0.516902 \\0.102977	0.519389 \\0.103974	0.521844 \\0.10617	0.527472 \\0.106781	0.529037 \\0.107574	0.531068 \\0.112181	0.542874 \\0.113587	0.546369 \\0.114778	0.54933 \\0.117791	0.556818 \\0.11818	0.557786 \\0.12299	0.569742 \\0.124186	0.57232 \\0.126597	0.577519 \\0.127988	0.580518 \\0.130398	0.585716 \\0.131391	0.587855 \\0.132197	0.589592 \\0.137202	0.600211 \\0.138195	0.602316 \\0.138801	0.603603 \\0.140397	0.607183 \\0.146801	0.621555 \\0.148809	0.626062 \\0.151811	0.632798 \\0.152204	0.633527 \\0.159015	0.646164 \\0.159409	0.646896 \\0.160616	0.649136 \\0.167814	0.66216 \\0.169022	0.664346 \\0.171216	0.668316 \\0.172223	0.670139 \\0.17662	0.680777 \\0.18043	0.689997 \\0.18083	0.690714 \\0.190235	0.707588 \\0.190836	0.708665 \\0.196833	0.7217 \\0.200643	0.72998 \\0.201044	0.730851 \\0.204839	0.7391 \\0.210442	0.751278 \\0.215453	0.762171 \\0.216847	0.765199 \\0.223651	0.779989 \\0.228054	0.789558 \\0.228062	0.789577 \\0.235058	0.804781 \\0.236459	0.807826 \\0.236668	0.808281 \\0.245664	0.827834 \\0.250868	0.839146 \\0.251679	0.840908 \\
    };
    
    \addplot+[smooth,mark=+, mark options={scale=1.5, solid},only marks] 
table[row sep=\\]{
 X Y\\
9E-05	0.033333 \\  	  0.01089	0.166667 \\  	 0.02709	0.233333 \\  	  0.03789	0.3 \\  0.05949	0.366667 \\  	  0.08109	0.433333 \\  0.10809	0.5 \\  0.13509	0.566667 	 \\  0.16749	0.633333 \\  0.20529	0.7 \\  	 0.23769	0.766667 \\  	  0.27549	0.833333	 \\  0.30789	0.9 	 \\  0.33489	0.966667 \\  	 
};

    \addplot+[smooth,mark=square, mark options={scale=1.5, solid},only marks] 
table[row sep=\\]{
 X Y\\
9E-05	0.033333 \\  	   0.01089	0.166667 \\  	 0.02709	0.233333 \\  	 0.03789	0.3 \\  0.05949	0.366667 \\   0.08109	0.433333 \\  	  0.10809	0.5 \\  0.13509	0.566667 \\  	  0.16749	0.633333 \\  	   0.20529	0.7 \\    0.23769	0.766667 \\  	 0.27549	0.833333 \\  	   0.30789	0.9 \\  	  0.33489	0.966667 \\  	 
};

    \addplot+[smooth,mark=o, mark options={scale=1.5, solid},only marks] 
table[row sep=\\]{
 X Y\\
 9E-05	0.033333 \\   0.02169	0.166667 \\   0.04329	0.3 	 \\  0.07569	0.366667	 \\  0.09189	0.433333   	 \\  0.12969	0.5 \\  0.15129	0.566667 \\  	   0.18909	0.633333 \\   0.22149	0.7 \\    0.25929	0.766667  \\  0.29169	0.833333	 \\  0.32949	0.9 \\    0.35109	0.966667 \\  
};

    \addplot+[smooth,mark=x, mark options={scale=1.5, solid},only marks] 
table[row sep=\\]{
 X Y\\
9E-05	0.033333 \\  0.02169	0.166667 \\  0.04329	0.3 \\  0.07569	0.366667 \\  0.09189	0.433333	 \\  0.12969	0.5 	 \\  0.15129	0.566667 \\	 0.18909	0.633333 \\  	   0.22149	0.7 \\  0.25929	0.766667 \\  	  0.29169	0.833333  \\  0.32949	0.9 	 \\  0.35649	0.966667	 \\  
};

    \addplot+[smooth,mark=-, color=green, mark options={scale=1.5, solid},only marks] 
table[row sep=\\]{
 X Y\\
9E-05	0.033333 \\   0.01629	0.166667 \\  0.03249	0.233333 \\  	 \\  0.04329	0.3 \\    0.06489	0.366667 \\  	  0.08649	0.433333 \\  0.10809	0.5 \\ 0.14049	0.566667 \\  	  0.17289	0.633333 	 \\  0.20529	0.7 \\  0.23769	0.766667 \\  	  0.28089	0.833333  \\  0.30249	0.9 \\  	 0.34029	0.966667 \\  	 
};

\legend{Hsiao \cite{Hsiao_86}~~~~~, Cao \cite{Cao_89}~~~~~~, Keung \cite{Keung_80}~~, Equation ~\ref{eq:Cp_20},Equation ~\ref{eq:Cp_21}, Equation \ref{eq:Cp_12},Equation \ref{eq:Cp_15},Equation \ref{eq:Cp_18}}
    \end{axis}
\end{tikzpicture}
\caption{Position of the melting front moving along the diagonal of a cuboid ($30 \times 30$)}
\label{fig:Cp_6}
\end{figure}

To assess mesh dependency, the previous problem is solved on a much coarser $30 \times 30$ mesh. Results are shown in figure \ref{fig:Cp_6} along with the same results of Hsiao et al. \cite{Hsiao_86}, Cao et al. \cite{Cao_89}, and Keung \cite{Keung_80} with a mesh resolution of $20 \times 20$. It is clear that by reducing the resolution, equations \ref{eq:Cp_12} and \ref{eq:Cp_15} start to deviate from the converged values, while equations \ref{eq:Cp_20}, \ref{eq:Cp_21}, and \ref{eq:Cp_18} show near to no sensitivity to this decrease in solution resolution. This shows how these equations are more accurate at coarser resolutions.

As figure \ref{fig:Cp_7} shows that equation \ref{eq:Cp_18} is the most accurate, the rest of this section is focused on further examination of the functionality of this equation. 

\begin{figure}[h]
        \centering
                \includegraphics[scale=0.8]{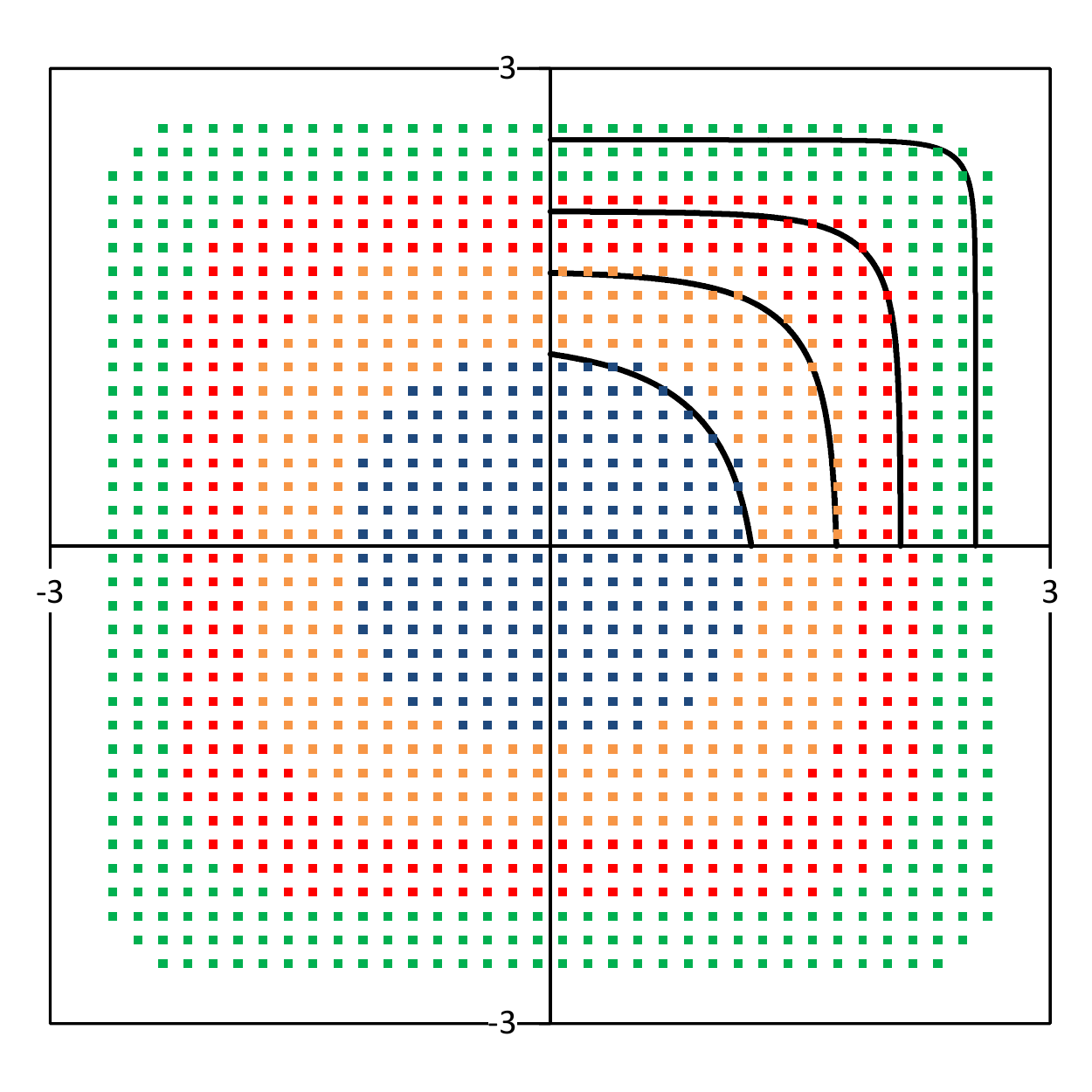}
\includegraphics[scale=0.8]{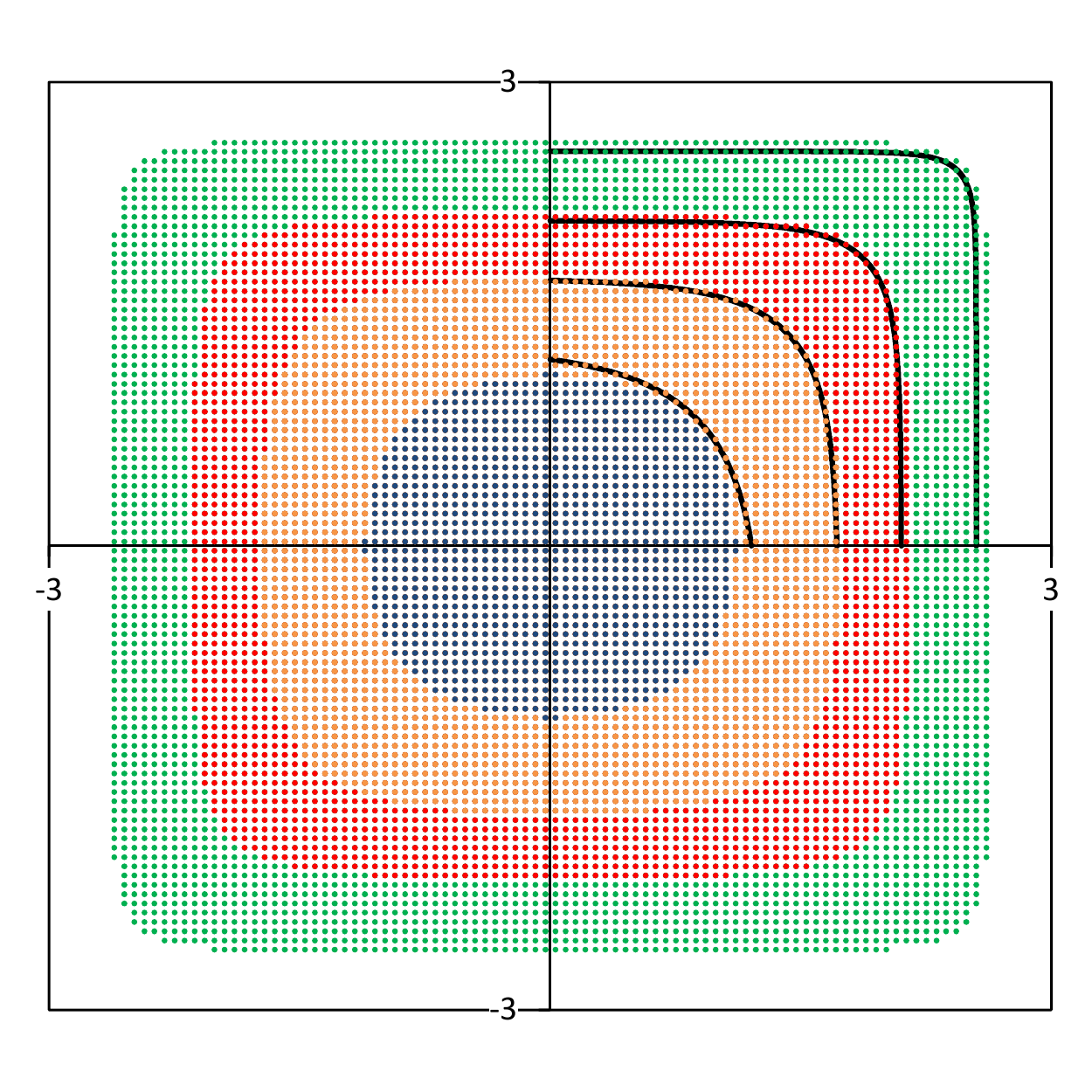}
\caption{Melting of a cuboid at $0.1$, $0.4$, $0.8$, $1.2$, and $1.44$ seconds. Solid lines are the analytical solution. Solution resolution: $40 \times 40$ (top) and $100 \times 100$ (bottom).}
\label{fig:Cp_8}
\end{figure}

Figure \ref{fig:Cp_8} illustrates the melting of a cuboid, for $\theta=0.3, St=4$, $\alpha_l/\alpha_s =1$, and $k_l/k_s=1$. Results are compared at two resolutions, $40 \times 40$ and $100 \times 100$, at $0.1$, $0.4$, $0.8$, $1.2$, and $1.44$ seconds (figure \ref{fig:Cp_8}). The solid lines in this figure show the analytical solution.

\subsubsection{Effect of $\Delta T$}

Equation \ref{eq:Cp_18} assumes that phase change occurs over a temperature range $\Delta T$, the value of which must be chosen for each problem. It was mentioned previously that the value of $\Delta T$ must be small enough to approximate physical reality, yet if $\Delta T$ is too small, some particles might not experience phase change as the temperature can jump from a value above (below) melting to a value below (above) in one time step during solidification (melting). This must be taken into consideration when choosing $\Delta T$ and $\Delta t$. For the previous test cases, a dimensionless value of $\Delta T=0.02$ was used.

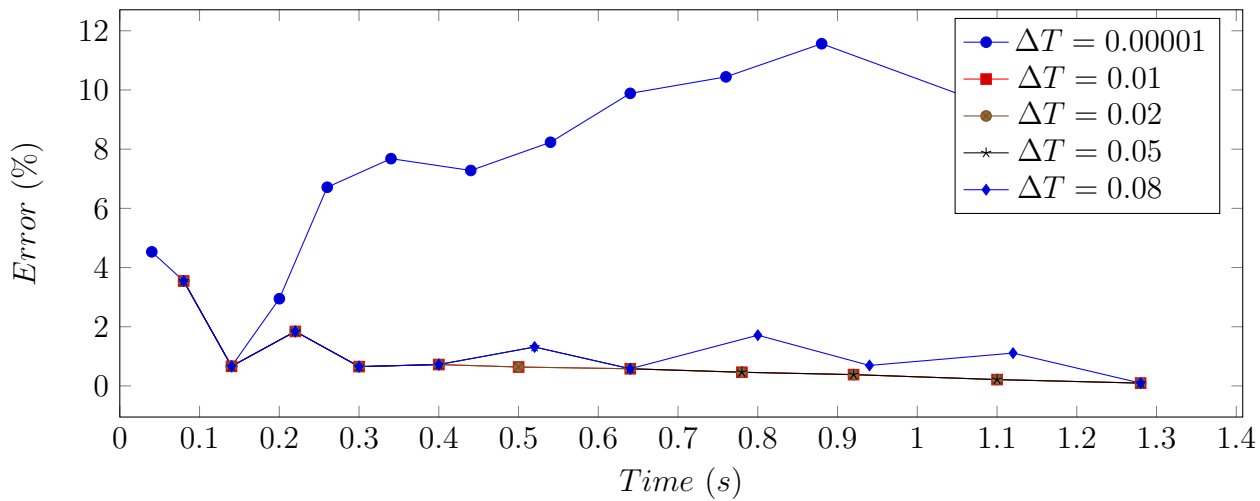
\begin{figure}[h]
        \centering
\begin{tikzpicture}
    \begin{axis}[xlabel=$Time~(s)$, width=\textwidth,height=7cm, ylabel=$Error~(\%)$, xmin=0]

    \addplot
table[row sep=\\]{
 X Y\\
0.0401	4.530317 \\ 0.0801	 3.544309 \\ 0.1401	0.662542 \\ 0.2001	2.946975 \\ 0.2601	6.713073 \\ 0.3401	7.679342 \\ 0.4401	7.279734 \\ 0.5401	8.233316 \\ 0.6401	9.885359 \\ 0.7601	10.44268 \\ 0.8801 	11.56255 \\ 1.0601	9.782925 \\
};

    \addplot
table[row sep=\\]{
 X Y\\
0.0801	3.544309 \\  0.1401	0.662542 \\  0.2201	1.84168 \\  0.3001	0.652989 \\  0.4001	0.722362 \\  0.5001	0.638683 \\  0.6401	0.579913 \\  0.7801	0.46207 \\  0.9201	0.381752 \\  1.1001	0.214284 \\  1.2801	0.09522 \\  
};

    \addplot
table[row sep=\\]{
 X Y\\
0.0801	3.544309 \\  0.1401	0.662542 \\  0.2201	1.84168 \\  0.3001	0.652989 \\  0.4001	0.722362 \\  0.5001	0.638683 \\  0.6401	0.579913 \\  0.7801	0.46207 \\  0.9201	0.381752 \\  1.1001	0.214284 \\  1.2801	0.09522 \\  
};

    \addplot
table[row sep=\\]{
 X Y\\
 0.0801	3.544309 \\  0.1401	0.662542 \\  0.2201	1.84168 \\  0.3001	0.652989 \\  0.4001	0.722362 \\  0.5201	1.315273 \\  0.6401	0.579913 \\  0.7801	0.46207 \\  0.9201	0.381752 \\  1.1001	0.214284 \\  1.2801	0.09522 \\  
};

    \addplot
table[row sep=\\]{
 X Y\\
0.0801	3.544309 \\  0.1401	0.662542 \\  0.2201	1.84168 \\  0.3001	0.652989 \\  0.4001	0.722362 \\  0.5201	1.315273 \\  0.6401	0.579913 \\  0.8001	1.714011 \\  0.9401	0.691766 \\  1.1201	1.109161 \\  1.2801	0.09522 \\  
};
\legend{$\Delta T=0.00001$,$\Delta T=0.01~~~~$,$\Delta T=0.02~~~~$,$\Delta T=0.05~~~~$,$\Delta T=0.08~~~~$}
    \end{axis}
\end{tikzpicture}
\caption{Error of the solidification front position versus the analytical solution, for solidification in a corner, for various $\Delta T$.}
\label{fig:Cp_9}
\end{figure}

The melting cuboid problem is considered again, with $T_i^*=0.3, St=4$, $\alpha_l/\alpha_s =1$, and $k_l/k_s =1$. $\Delta T$ is set to values of $0.00001$, $0.01$, $0.02$, $0.05$, and $0.08$. The error values for this test case, compared to the analytical solution, are plotted in figure \ref{fig:Cp_9}. For very small $\Delta T$ ($0.00001$ here), errors are large as many particles do not experience the latent heat, and at large $\Delta T$ ($0.08$ here), the error grows as the phase change temperature interval deviates too far from the physical reality. The results of various test cases, including the one reported here, confirm that there exists a large range of $\Delta T$ over which the results are relatively independent of the choice of $\Delta T$.

\subsubsection{Effect of Smoothing Kernel}

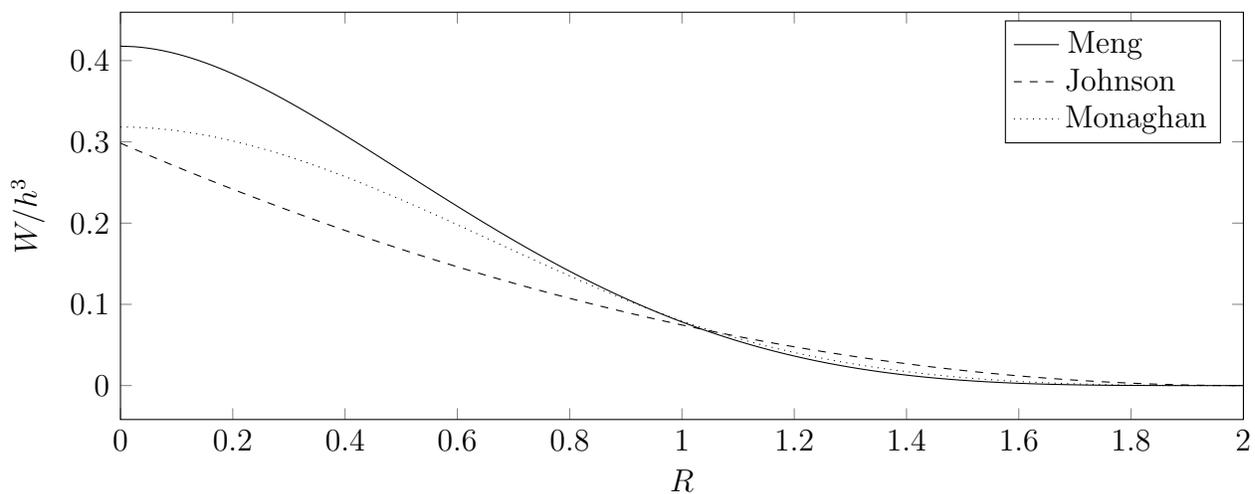
\begin{figure}[h]
        \centering
\begin{tikzpicture}
    \begin{axis}[xlabel=$ R$, width=\textwidth,height=7cm, ylabel=$W / h^3$,cycle list name=black white,xmin=0,xmax=2]

    \addplot+[smooth,mark=none] 
table[row sep=\\]{
 X Y\\
0	0.417781726 \\  0.025	0.417145109 \\  0.05	0.415298715 \\  0.075	0.41233503 \\  0.1	0.408342992 \\  0.125	0.403408048 \\  0.15	0.397612219 \\  0.175	0.391034162 \\  0.2	0.383749226 \\  0.225	0.375829519 \\  0.25	0.367343966 \\  0.275	0.35835837 \\  0.3	0.348935475 \\  0.325	0.339135026 \\  0.35	0.329013831 \\  0.375	0.31862582 \\  0.4	0.308022111 \\  0.425	0.297251064 \\  0.45	0.286358349 \\  0.475	0.275387005 \\  0.5	0.264377498 \\  0.525	0.253367788 \\  0.55	0.242393383 \\  0.575	0.231487409 \\  0.6	0.220680663 \\  0.625	0.210001679 \\  0.65	0.199476788 \\  0.675	0.189130178 \\  0.7	0.178983958 \\  0.725	0.169058215 \\  0.75	0.159371081 \\  0.775	0.149938787 \\  0.8	0.14077573 \\  0.825	0.131894534 \\  0.85	0.123306106 \\  0.875	0.115019703 \\  0.9	0.107042989 \\  0.925	0.099382101 \\  0.95	0.092041704 \\  0.975	0.085025057 \\  1	0.078334074 \\  1.025	0.07196938 \\  1.05	0.065930379 \\  1.075	0.060215312 \\  1.1	0.054821318 \\  1.125	0.049744495 \\  1.15	0.044979964 \\  1.175	0.040521924 \\  1.2	0.036363721 \\  1.225	0.032497905 \\  1.25	0.028916289 \\  1.275	0.025610015 \\  1.3	0.022569613 \\  1.325	0.019785062 \\  1.35	0.01724585 \\  1.375	0.014941039 \\  1.4	0.012859322 \\  1.425	0.010989086 \\  1.45	0.009318475 \\  1.475	0.007835447 \\  1.5	0.006527839 \\  1.525	0.005383428 \\  1.55	0.004389988 \\  1.575	0.003535357 \\  1.6	0.002807493 \\  1.625	0.00219454 \\  1.65	0.001684884 \\  1.675	0.00126722 \\  1.7	0.000930609 \\  1.725	0.000664539 \\  1.75	0.000458989 \\  1.775	0.000304489 \\  1.8	0.00019218 \\  1.825	0.000113877 \\  1.85	6.21287E-05 \\  1.875	3.02805E-05 \\  1.9	1.25335E-05 \\  1.925	4.00697E-06 \\  1.95	7.9966E-07 \\  1.975	5.04888E-08 \\  2	0 \\  
};

    \addplot+[smooth,mark=none,dashed] 
table[row sep=\\]{
 X Y\\
0	0.298415518 \\  0.025	0.291001758 \\  0.05	0.283681252 \\  0.075	0.276454001 \\  0.1	0.269320005 \\  0.125	0.262279264 \\  0.15	0.255331778 \\  0.175	0.248477546 \\  0.2	0.24171657 \\  0.225	0.235048848 \\  0.25	0.228474381 \\  0.275	0.221993169 \\  0.3	0.215605212 \\  0.325	0.20931051 \\  0.35	0.203109062 \\  0.375	0.19700087 \\  0.4	0.190985932 \\  0.425	0.185064249 \\  0.45	0.179235821 \\  0.475	0.173500647 \\  0.5	0.167858729 \\  0.525	0.162310065 \\  0.55	0.156854657 \\  0.575	0.151492503 \\  0.6	0.146223604 \\  0.625	0.14104796 \\  0.65	0.135965571 \\  0.675	0.130976436 \\  0.7	0.126080556 \\  0.725	0.121277932 \\  0.75	0.116568562 \\  0.775	0.111952447 \\  0.8	0.107429587 \\  0.825	0.102999981 \\  0.85	0.098663631 \\  0.875	0.094420535 \\  0.9	0.090270694 \\  0.925	0.086214108 \\  0.95	0.082250777 \\  0.975	0.078380701 \\  1	0.07460388 \\  1.025	0.070920313 \\  1.05	0.067330001 \\  1.075	0.063832944 \\  1.1	0.060429142 \\  1.125	0.057118595 \\  1.15	0.053901303 \\  1.175	0.050777266 \\  1.2	0.047746483 \\  1.225	0.044808955 \\  1.25	0.041964682 \\  1.275	0.039213664 \\  1.3	0.036555901 \\  1.325	0.033991393 \\  1.35	0.031520139 \\  1.375	0.02914214 \\  1.4	0.026857397 \\  1.425	0.024665908 \\  1.45	0.022567674 \\  1.475	0.020562694 \\  1.5	0.01865097 \\  1.525	0.0168325 \\  1.55	0.015107286 \\  1.575	0.013475326 \\  1.6	0.011936621 \\  1.625	0.010491171 \\  1.65	0.009138975 \\  1.675	0.007880035 \\  1.7	0.006714349 \\  1.725	0.005641918 \\  1.75	0.004662742 \\  1.775	0.003776821 \\  1.8	0.002984155 \\  1.825	0.002284744 \\  1.85	0.001678587 \\  1.875	0.001165686 \\  1.9	0.000746039 \\  1.925	0.000419647 \\  1.95	0.00018651 \\  1.975	4.66274E-05 \\  2	0 \\  
};

    \addplot+[smooth,mark=none,dotted] 
table[row sep=\\]{
 X Y\\
0	0.318309886 \\  0.025	0.318015201 \\  0.05	0.317146066 \\  0.075	0.315724862 \\  0.1	0.31377397 \\  0.125	0.311315772 \\  0.15	0.308372649 \\  0.175	0.304966982 \\  0.2	0.301121152 \\  0.225	0.296857541 \\  0.25	0.292198528 \\  0.275	0.287166497 \\  0.3	0.281783827 \\  0.325	0.2760729 \\  0.35	0.270056097 \\  0.375	0.263755799 \\  0.4	0.257194388 \\  0.425	0.250394244 \\  0.45	0.24337775 \\  0.475	0.236167285 \\  0.5	0.228785231 \\  0.525	0.221253969 \\  0.55	0.213595881 \\  0.575	0.205833347 \\  0.6	0.197988749 \\  0.625	0.190084468 \\  0.65	0.182142885 \\  0.675	0.174186381 \\  0.7	0.166237338 \\  0.725	0.158318136 \\  0.75	0.150451157 \\  0.775	0.142658782 \\  0.8	0.134963392 \\  0.825	0.127387368 \\  0.85	0.119953091 \\  0.875	0.112682943 \\  0.9	0.105599305 \\  0.925	0.098724557 \\  0.95	0.092081082 \\  0.975	0.085691259 \\  1	0.079577472 \\  1.025	0.073757126 \\  1.05	0.068227735 \\  1.075	0.062981839 \\  1.1	0.058011977 \\  1.125	0.053310689 \\  1.15	0.048870515 \\  1.175	0.044683994 \\  1.2	0.040743665 \\  1.225	0.03704207 \\  1.25	0.033571746 \\  1.275	0.030325234 \\  1.3	0.027295073 \\  1.325	0.024473803 \\  1.35	0.021853963 \\  1.375	0.019428094 \\  1.4	0.017188734 \\  1.425	0.015128423 \\  1.45	0.013239702 \\  1.475	0.011515109 \\  1.5	0.009947184 \\  1.525	0.008528467 \\  1.55	0.007251497 \\  1.575	0.006108814 \\  1.6	0.005092958 \\  1.625	0.004196468 \\  1.65	0.003411884 \\  1.675	0.002731745 \\  1.7	0.002148592 \\  1.725	0.001654963 \\  1.75	0.001243398 \\  1.775	0.000906437 \\  1.8	0.00063662 \\  1.825	0.000426486 \\  1.85	0.000268574 \\  1.875	0.000155425 \\  1.9	7.95775E-05 \\  1.925	3.35717E-05 \\  1.95	9.94718E-06 \\  1.975	1.2434E-06 \\  2	0 \\  
};
\legend{Meng~~~~~~~, Johnson~~~~, Monaghan}
    \end{axis}
\end{tikzpicture}
\caption{$W(R,h)/h^3$ for the three smoothing kernels.}
\label{fig:Cp_13}
\end{figure}

Finally, the choice of smoothing kernel that appears in equation \ref{eq:Cp_17} is important as it dictates how the latent heat is released into the domain. Table \ref{tab:Cp_1} lists three smoothing functions (from \cite{Meng_14,Johnson_96,Monaghan_85}). These kernels are plotted in figure \ref{fig:Cp_13}. To compare these kernels, $W^T$ and $W^P$ are taken to be identical, where the former is in the 1D domain of temperature and the latter is in the 3D $x-y-z$ domain. This only changes the constant coefficients of these kernels. For instance, the Meng \cite{Meng_14} kernel in 3D has a coefficient of $21/(16 \pi h^3 )$ for $W^P$ and a coefficient of  $3/4h$ (in 1D) for $W^T$. Note that the $\alpha ^{3D}$ values are used for the 2D test case here, since it is solved in a 3D domain.

\begin{table}[h]
\caption[Smoothing kernels]{Smoothing kernels}
\label{tab:Cp_1}
\begin{tabular*}{\textwidth}{c @{\extracolsep{\fill}} ccccc}
 \hline
\\
Kernel  &  $\alpha^{3D}$  &  $\alpha^{1D}$ &  Ref\\\\
 \hline
$
 W(R,h) =\alpha \left\{ \begin{array}{ll}
( 1-R/2)^4(2R+1) & \mbox{$0 \leq R \leq 2$}\\
0 & \mbox{$otherwise$}\\
\end{array} \right. 
$ 
&  $\frac{21}{16\pi h^3}$  & $\frac{3}{4 h}$ & \cite{Meng_14}
\\\\ 
$
 W(R,h) =\alpha \left\{ \begin{array}{ll}
(3/16)R^2-(3/4)R+3/4 & \mbox{$0 \leq R \leq 2$}\\
0 & \mbox{$otherwise$}\\
\end{array} \right. 
$ 
&  $\frac{5}{4\pi h^3}$  & $\frac{1}{h}$ & \cite{Johnson_96}
\\\\
$
 W(R,h) =\alpha \left\{ \begin{array}{ll}
2/3-R^2+R^3/2 & \mbox{$0 \leq R < 1$}\\
(1/6)(2-R)^3 & \mbox{$1 \leq R < 2$}\\
0 & \mbox{$otherwise$}\\
\end{array} \right. 
$  
&  $\frac{3}{2\pi h^3}$  & $\frac{1}{h}$ & \cite{ Monaghan_85}
\\
 \hline
\end{tabular*}
\end{table}

\begin{figure}[h]
        \centering
\begin{tikzpicture}
    \begin{axis}[xlabel=$Time~(s)$, width=\textwidth,height=7cm, ylabel=$Error~(\%)$,cycle list name=black white, ymax=10]

    \addplot+[mark=none] 
table[row sep=\\]{
 X Y\\
0.1001	7.375590524\\ 0.1401	0.662541857\\ 0.2201	1.841680446\\ 0.3201	3.80666849\\ 0.4001	0.722362385\\ 0.5401	3.159664889\\ 0.6401	0.579913137 \\  0.8001	1.71401138\\ 0.9401	0.691765618 \\  1.1201	1.109161128 \\  1.2801	0.095220351\\ 
};
    \addplot+[mark=none,dashed] 
table[row sep=\\]{
 X Y\\
0.0801	3.54430868 \\  0.1401	0.662541857 \\  0.2001	2.946975456 \\  0.2801	2.832702342 \\0.4001	0.722362385 \\   0.4801	2.713497888	 \\  0.6201	1.010654741  \\  0.7601	0.838964972 \\   0.9001	1.490853799\\  1.0601	1.650856499 \\  1.2201	2.331773638 \\  1.3801	3.344452991 \\ 
};
    \addplot+[mark=none,dotted] 
table[row sep=\\]{
 X Y\\
 0.0801	3.54430868 \\  0.1401	0.662541857 \\  	 0.1801	8.51262753  \\  0.2801	2.832702342 \\  0.3601	4.64636548 \\  0.4801	2.713497888 \\  0.5801	4.435131786 \\  	 0.7201	3.601809342 \\    0.8601	3.82401345 \\  1.0001	4.655666036 \\    1.1601	4.944697889 	 \\  1.3201	5.666916242 \\ 
};
\legend{Meng~~~~~~~, Johnson~~~~, Monaghan}
    \end{axis}
\end{tikzpicture}
\caption{Error of the solidification front position versus the analytical solution, for three different kernels.}
\label{fig:Cp_14}
\end{figure}
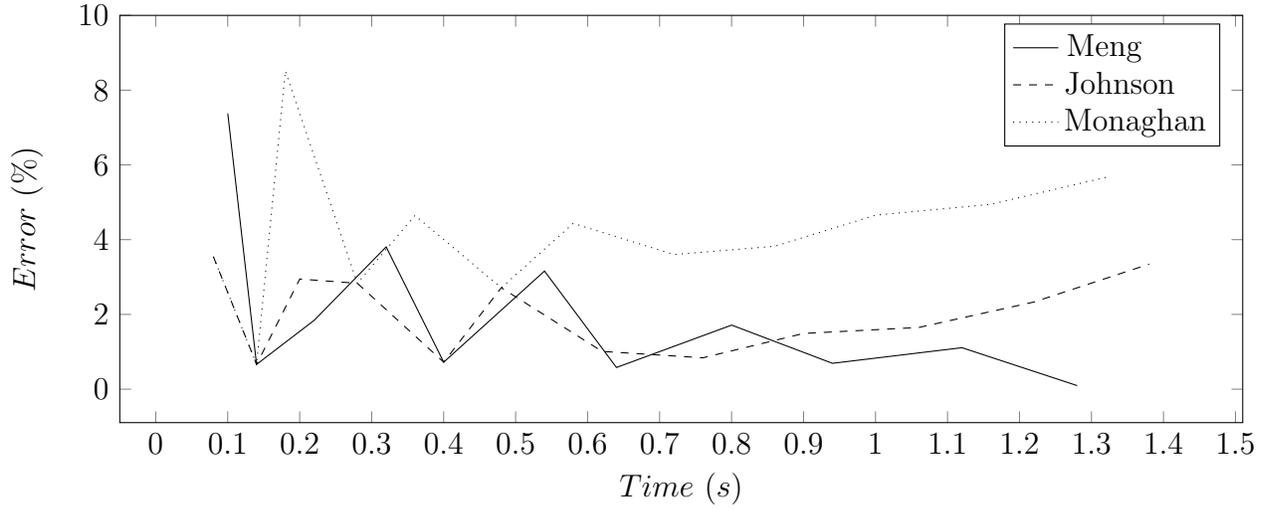

For the case of the solidification of the cuboid, results using these three kernels are compared to the analytical solution. Figure \ref{fig:Cp_14} shows that all three kernels produce accurate results with an error below 5\%, but the Meng et al. \cite{Meng_14} and Johnson et al. \cite{Johnson_96} kernels produce somewhat more accurate solutions.

\subsection{Three-Dimensional Problems}

In this section, the validity of the model of equation \ref{eq:Cp_18} is assessed on two 3D test cases.

\subsubsection{Melting of a sphere}

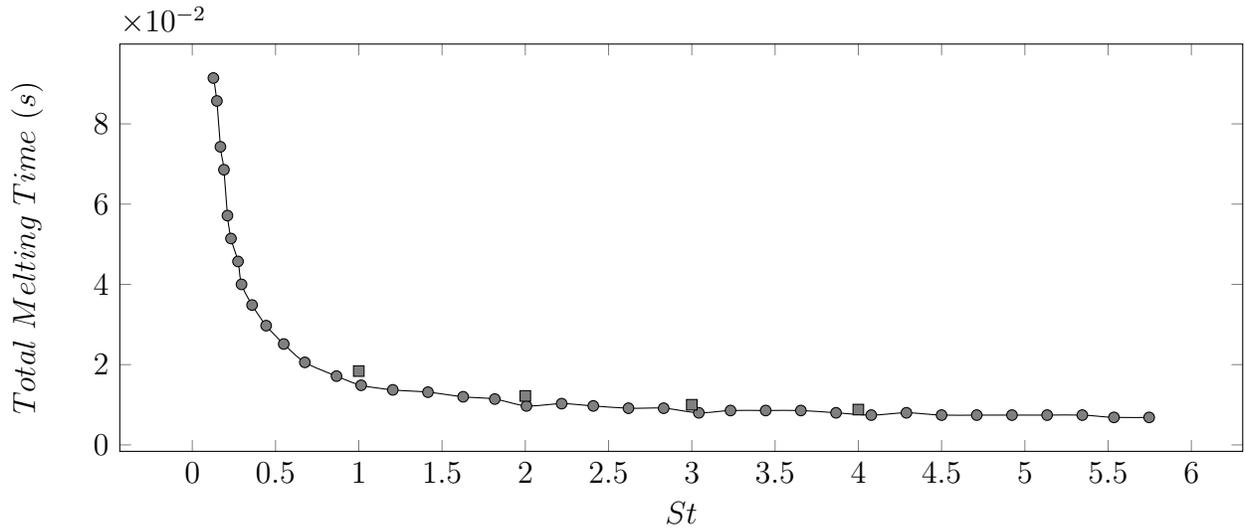
\begin{figure}[h]
        \centering
\begin{tikzpicture}

    \begin{axis}[xlabel=$St$, width=\textwidth, height=7cm, ylabel=$Total~Melting~Time~(s)$,cycle list name=black white,/pgfplots/tick scale binop=\times]

    \addplot+[smooth] 
table[row sep=\\]{
 X Y\\
0.126761	0.0914286 \\  0.147887	0.0857143 \\  0.169014	0.0742857 \\  0.190141	0.0685714 \\  0.211268	0.0571429 \\  0.232394	0.0514286 \\  0.274648	0.0457143 \\  0.295775	0.04 \\  0.359155	0.0348571 \\  0.443662	0.0297143 \\  0.549296	0.0251429 \\  0.676056	0.0205714 \\  0.866197	0.0171429 \\  1.01408	0.0148571 \\  1.20423	0.0137143 \\  1.41549	0.0131429 \\  1.62676	0.012 \\  1.8169	0.0114286 \\  2.00704	0.00971429 \\  2.21831	0.0102857 \\  2.40845	0.00971429 \\  2.61972	0.00914286 \\  2.83099	0.00914286 \\  3.04225	0.008 \\  3.23239	0.00857143 \\  3.44366	0.00857143 \\  3.65493	0.00857143 \\  3.8662	0.008 \\  4.07746	0.00742857 \\  4.28873	0.008 \\  4.5	0.00742857 \\  4.71127	0.00742857 \\  4.92254	0.00742857 \\  5.1338	0.00742857 \\  5.34507	0.00742857 \\  5.53521	0.00685714 \\  5.74648	0.00685714 \\  
};

    \addplot+[smooth,only marks] 
table[row sep=\\]{
 X Y\\
1 0.0184\\
2 0.0122\\
3 0.01\\
4 0.0088\\
};

    \end{axis}
\end{tikzpicture}
\caption{Melting of a sphere. Total time of melting versus Stefan number. Squares show results obtained using the model presented here, and the solid line is from the mathematical model of \cite{Alexiades_92}.}
\label{fig:Cp_10}
\end{figure}

The first 3D test case is the melting of a sphere. Problem specifications are the same as in \cite{Ayasoufi_04}. A sphere is considered to be initially at its melting temperature ($T_i=0^\circ C$). At time 0, the temperature of the wall surrounding the sphere is raised to a value higher than the melting temperature ($T_w=1^\circ C$). In this test case, similar to the test performed by \cite{Ayasoufi_04}, the Stefan number $St$ is varied from 1 to 4. The sphere has a radius of $0.25m$ and specific heats and thermal diffusivities are set to unity. Figure \ref{fig:Cp_10} shows the total melting time for the sphere at different Stefan numbers. The melting times obtained using equation \ref{eq:Cp_18}, as can be seen in figure \ref{fig:Cp_10}, are in good agreement with the mathematical model of \cite{Alexiades_92}.

\begin{figure}[h]
        \centering
                \includegraphics{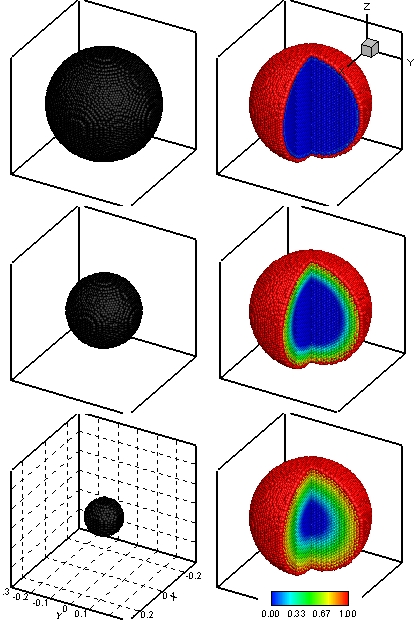}
\caption{Melting of a sphere of radius $0.25$, with a wall temperature $T_w=1^\circ C$, at $t=0, 0.0025$, and $0.0065$ seconds. Left: un-melted particles; Right: temperature.}
\label{fig:Cp_11}
\end{figure}

Particle temperatures for the case of $St=4$ are plotted in figure \ref{fig:Cp_11} at $t=0, 0.0025$, and $0.0065$.

\subsubsection{Melting of a cone}

\begin{figure}[h]
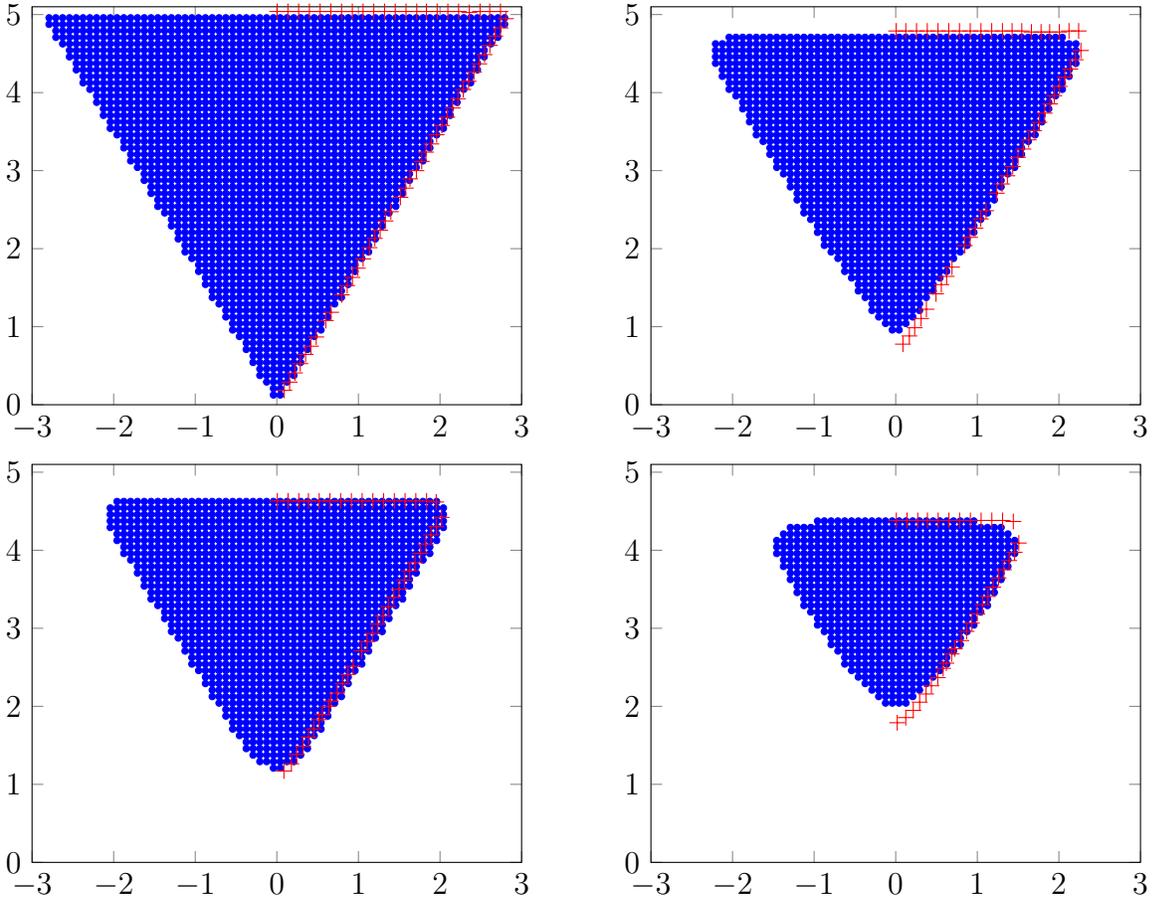

        \centering
                \begin{subfigure}[b]{0.49\textwidth}

        \end{subfigure}
\caption{Melting of a cone. Interface position is plotted at different times: 0.05, 0.1, 0.25, and 0.5. Red crosses are the results of Ayasoufi et al. \cite{Ayasoufi_03} obtained using a space-time conservation element and solution element (CE/SE) method.}
\label{fig:Cp_12}
\end{figure}

The second 3D test is the melting of a cone. For comparison, the problem specifications have been chosen similar to a test case from \cite{Ayasoufi_03}. A cone with a half angle of 30 degrees is initially at $T_i=0^\circ C$. The temperature on the wall surrounding the cone is increased to $T_w=0.3^\circ C$ at $t=0$. The phase change interface at different times is shown in figure \ref{fig:Cp_12}, and compared to the results of Ayasoufi et al. \cite{Ayasoufi_03} (CE/SE method). As can be seen, the SPH solution yields results close to those of \cite{Ayasoufi_03}, while slightly underpredicting the melt front.

\subsection{Computational Cost}
Finally, the SPH solver used here was developed to run entirely on graphics processing units (GPU). The device used for this study is the NVIDIA\textsuperscript{\textregistered} Tesla\textsuperscript{\textregistered} K20 GPU accelerator with 2496 processor cores and a processor core clock of 706 MHz. For the sample conduction problems with phase change, 10.3 kB of memory per particle was needed on the GPU device. The five equations mentioned before perform quite consistently with regard to the amount of computational time, except for equation \ref{eq:Cp_12}, which is near 8\% faster than the rest. This is because equation \ref{eq:Cp_12} approximates the amount of latent heat released (absorbed) without the need of smoothing steps or evaluating gradients. However, equation \ref{eq:Cp_12} is also the least accurate. Of the rest of the equations, equation \ref{eq:Cp_18} requires the least number of computation steps, and so is easier to implement.